\documentclass[aps,superscriptaddress,showkeys,showpacs,longbibliography,nofootinbib]{revtex4-2}
\usepackage{graphicx}
\usepackage{bm}
\usepackage{graphicx}
\usepackage{amsmath}
\usepackage{mathrsfs}
\usepackage{ulem}
\usepackage{color}
\usepackage[colorlinks, linkcolor=red,anchorcolor=green,citecolor=blue]{hyperref}

\newcommand{\nn}{\nonumber}
\newcommand\diag{\operatorname{diag}}

\graphicspath{{./figs/}}

\begin{document}  
	
\title{Linear mode analysis from spin transport equation}

\author{Jin Hu}
\email{hu-j17@mails.tsinghua.edu.cn}
\affiliation{Department of Physics, Tsinghua University, Beijing 100084, China}

\begin{abstract}
We provide a linear analysis on normal modes of the spin Boltzmann equation proposed in \cite{Weickgenannt:2021cuo}, where the non-diagonal or polarized part of the transition rate is neglected to ensure the Hermitian property of linearized collision operator. As an instrumental element of spin kinetic theory,  the  conservation of total angular momentum is explicitly considered,  thus our analysis is relevant to the recent investigation on the issue of  local spin polarization. By treating the linearized collision operator as an evolution operator,  solving the normal modes turns out a degenerate perturbation problem in quantum mechanics.  The dispersion relations of spinless modes  are in accordance with well-known calculations,  while  the frequencies of spin modes  are also determined  up to second-order in wave vector and  the second order expressions  are only formal solutions to be further determined. Moreover, the relaxation of spin density is related to our linear mode analysis, which shall play a big role in investigating the issues of  the local spin polarization in the relativistic heavy-ion collisions.
\end{abstract}




\maketitle	
\section{Introduction}
The community of heavy-ion collisions shows great interests towards the research on spin polarization  inspired by the
recent measurements of  the spin-related observables of $ \Lambda $ hyperons \cite{STAR:2017ckg,Alpatov:2020iev}.   Although satisfying agreements with experiment data in global polarization are reported \cite{Wei:2018zfb,Karpenko:2016jyx,Csernai:2018yok,Li:2017slc,Bzdak:2017shg,Shi:2017wpk,Sun:2017xhx,Ivanov:2019wzg,Xie:2017upb},  the ``spin sign puzzle'' long bothers the scientists,   where the dependence of $ \Lambda $ polarization  on the azimuthal angle and transverse momentum \cite{Adam:2019srw,Adam:2018ivw} can not be well-reproduced  and  even the opposite dependence is predicted \cite{Becattini:2017gcx,Xia:2018tes}. 
It is widely believed that the origin of spin sign problem may be the inappropriate application of the equilibrium picture of spin extensively adopted in those numerical calculations.  To overcome it, the framework needs to take  non-equilibrium effects into consideration and spin hydrodynamic theory is thought to be a promising one. Compared to  ordinary hydrodynamics, spin hydrodynamics includes the conservation law of total angular momentum and focus on the dissipation of spin density. On the other hand, these direct experimental measurements of quantum effects in relativistic heavy-ion collisions provide the opportunity to study the evolution of spinful fluids and stimulate the related researches on spin kinetic theory.

Many efforts are along the line of  spin hydrodynamics trying to get insight into the spin sign problem \cite{Florkowski:2017ruc,Peng:2021ago,Becattini:2009wh, Florkowski:2018fap,Hattori:2019lfp,Bhadury:2020cop,Fukushima:2020ucl,Hu:2021lnx,Hu:2021pwh,Hu:2022xjn,Hu:2022mvl}.  Though with great progress in investigating spinful fluids, it is noted that the ultimate goal for us is to construct a causal and numerically stable theory of spin hydrodynamics, which allows the numerical implementation and simulation of the evolution of the fluid system, and eventually provides us with quantitative explanations for the spin-related experimental phenomena. To that end, the hydrodynamic equations and relevant transport coefficients must be both  obtained  from the microscopic spin kinetic theory, constructing a self-consistent theory.  Therefore it is necessary to derive the quantum transport equations with proper collision terms, which must incorporate spin as the independent variable and account for the coupling between spin and orbit. To note,  the related developments in spin transport can be found in \cite{Weickgenannt:2021cuo,Weickgenannt:2020aaf,Yang:2020hri,Wang:2020pej,Sheng:2021kfc}. Accounting for the complexity and non-linearity, these transport equations are hard to solve and the linearization of related transport equations are usually used as a procedure to get some analytical results or spin hydrodynamics, from which the low energy effective theory of spinful fluids can be constructed on solid ground. Following this manner, a second-order spin hydrodynamic theory \cite{Weickgenannt:2022zxs} is constructed from spin Boltzmann equation \cite{Weickgenannt:2021cuo,Weickgenannt:2020aaf} recently, while the causality and stability of resulting theory need further investigation.

In this work, we want to give a detailed linear analysis on normal modes appearing in spin hydrodynamic theory. To achieve it, the framework constructed in \cite{Weickgenannt:2021cuo} is adopted as the start point of our derivation. Although we are not directly  constructing  the hydrodynamic theory throughout the paper,  the theory of hydrodynamics can be completely constructed from these normal modes. After the linearization of the collision term, solving the linear integral equation can be ascribed to the solution to the spectrum of the linear collision operator while the non-equilibrium distribution can be expanded with  corresponding eigenfunctions. However, the full spectrum is not yet solved even for the local Boltzmann collision term. As will be shown around  Eq.(\ref{t1}), most modes decay rapidly except the zero modes protected by all conservation laws respected by the collision term.
Considering that the hydrodynamics actually originates from the conservation laws, 
the normal modes to be solved are equivalent to the description of spin hydrodynamics in this way.
This paper is organized as follows. In Sec.~\ref{secqu}
we  review the spin Boltzmann equation \cite{Weickgenannt:2021cuo}. In Sec.~\ref{linear} we linearize the nonlocal collision kernel and neglect the polarized transition rate to get the Hermitian collision operator.  In Sec.~\ref{expansion},  the dispersion relations of normal modes are determined following the method of degenerate perturbation theory \cite{Balescu, Resi}.
 In Sec.~\ref{relax}, we relate our results obtained in Sec.~\ref{expansion} to the relaxation of spin, which is relevant in the investigation of local spin polarization and thus draws much attention. Discussions and outlook are put in Sec.~\ref{su}. We use natural units $k_B=c=\hbar=1$. The metric tensor  is defined as  $g^{\mu\nu}\equiv\diag(1,-1,-1,-1)$ , while  the projection tensor orthogonal to  fluid velocity $u^\mu$ is given by $\Delta^{\mu\nu} \equiv g^{\mu\nu}-u^\mu u^\nu$.

In the following,  the shorthand notations are used:
\begin{eqnarray}
A^{( \mu\nu ) } &\equiv& (A^{ \mu\nu } + A^{ \nu \mu})/2, \\
A^{[ \mu\nu ] } &\equiv& (A^{ \mu\nu } - A^{ \nu \mu})/2, \\
A^{\langle \mu\nu \rangle}&\equiv&
	\bigg(\frac{\Delta^{\mu}_{\alpha} \Delta^{\nu}_{\beta} 
		+ \Delta^{\nu}_{\alpha} \Delta^{\mu}_{\beta}}{2}
	- \frac{\Delta^{\mu\nu} \Delta_{\alpha\beta}}{3}\bigg)A^{\alpha\beta},
\end{eqnarray}
and we decompose the derivative $\partial$ as 
\begin{align}
\partial^\mu=u^\mu D+\nabla^\mu,\quad D\equiv u^\mu\partial_\mu,\quad \nabla^\mu\equiv\Delta^{\mu\nu}\partial_\nu.
\end{align}

\section{Review of the spin Boltzmann equation}
\label{secqu}	
In this section, we review 
the spin Boltzmann equation   derived in \cite{Weickgenannt:2021cuo}, which provides a description of the evolution of the system composed of  massive fermions,
 
\begin{align}
\label{boltz}
p\cdot \partial f(x,p,\bm{s})=C[f]+C_{s}[f] ,
\end{align}	 
where  $C[f]$ is  the collision kernel in which particles involved  change their momentum and spin while in $C_s[f]$ particles only exchange spin without momentum transfer,
\begin{eqnarray}
\label{cf}
C[f] &  \equiv& \int d\Gamma_1 d\Gamma_2 d\Gamma^\prime\,    
\mathcal{W}\,  
[f(x+\Delta_1,p_1,\bm{s}_1)f(x+\Delta_2,p_2,\bm{s}_2)
-f(x+\Delta,p,\bm{s})f(x+\Delta^\prime,p^\prime,\bm{s}^\prime)],\\
C_{s}[f] &\equiv & \int  d \Gamma_2\, dS_1(p)\,
\mathcal{D}\, f(x+\Delta_1,p,\bm{s}_1)f(x+\Delta_2,p_2,\bm{s}_2)\;, \label{Cs}
\end{eqnarray}
with
\begin{align}
 d\Gamma &\equiv d^4p\, \delta(p^2 - m^2)dS(p) ,\\
   \int dS(p)  &\equiv \sqrt{\frac{p^2}{3 \pi^2}} \int d^4\bm{s}\, \delta(\bm{s}\cdot\bm{s}+3)
   \delta(p\cdot \bm{s})\;.
 \end{align}
 Note that the phase space spanned by particle position $x$ and momentum $p$ is now extended to include the variable $\bm{s}$ as a classical description of spin  \cite{Zamanian:2010zz,Ekman:2017kxi,Ekman:2019vrv,Florkowski:2018fap,Bhadury:2020puc,Weickgenannt:2020aaf}.
 Here the spatial shift $\Delta$ is defined as 
 \begin{equation}
 \label{deltanon}
 \Delta^\mu\equiv -\frac{1}{2m(p\cdot\hat{t}+m)}\, 
 \epsilon^{\mu\nu\alpha\beta}p_\nu \hat{t}_\alpha \bm{s}_{\beta}\;.
 \end{equation}
 Additionally, the transition rates are  shown by
\begin{eqnarray}
\mathcal{W}&\equiv& \delta^{(4)}(p+p^\prime-p_1-p_2)
\frac{1}{8} \sum_{s,r,s',r',s_1,s_2,r_1,r_2} h_{s r} (p,\bm{s})h_{s^\prime r^\prime}(p^\prime, \bm{s}^\prime) \,  
h_{s_1 r_1}(p_1, \bm{s}_1)\, h_{s_2 r_2}(p_2, \bm{s}_2) \nn \\
&&\times \langle{p,p^\prime;r,r^\prime|t|p_1,p_2;s_1,s_2}\rangle
\langle{p_1,p_2;r_1,r_2|t^\dagger|p,p^\prime;s,s^\prime}\rangle \;
\label{local_col_GLW_after}\\
\mathcal{D}&\equiv& \frac{\pi\hbar}{4m} \sum_{s_1,s_2,r,r_2}
\epsilon_{\mu\nu\alpha\beta} \bm{s}^\mu \bm{s}_1^{\nu} p^\alpha
n_{s_1r}^{\beta}(p)\, h_{s_2r_2} (p_2, \bm{s}_2)\langle{p,p_2;r,r_2|t+t^\dagger|p,p_2;s_1,s_2}\rangle \;,\nonumber\label{mathw}
\end{eqnarray}
with 
\begin{align}
\label{nsr}
n^\alpha_{s r}(p,\bm{s}) \equiv &\frac{1}{2m}\, \bar{u}_s(p)\gamma^5\gamma^\alpha u_r(p)\; ,\\
\label{hsr}
h_{s r}(p,\bm{s}) \equiv &\delta_{s r}+  \frac{1}{2m}\, \bar{u}_s(p)\gamma^5\bm{s} \cdot\gamma u_r(p)\; ,
\end{align} 
where $\gamma$ is Dirac matrix,  $u_s(p)$ is the spinor,  $r,s$ are spin indices, and the matrix element of $t$ is defined as the conventional scattering amplitude.

 In the following derivation, we choose to neglect $C_{s}[f]$. The reason for neglecting the spin-exchange term can be argued via an estimation of magnitude.
 When comparing $C_{s}[f]$ with $C[f]$, we find that the integration over $\bm{s}_1$ in $C_{s}[f]$ is not zero unless $f(x+\Delta_1,p,\bm{s}_1)$ can provide another $\bm{s}_1$ because of $	\int dS(p)\, \bm{s}^\mu  =  0$. The sources of $\bm{s}_1$ coming from $f(x+\Delta_1,p,\bm{s}_1)$ consist of two contributions. One is $\Delta_1$ accompanied with a derivative $\partial$, the other is $\Sigma_{\bm{s}_1}^{\mu\nu}$ which we expect to appear with $\Omega_{\mu\nu}$. In the following sections our discussion is limited to small $\Omega$ thus $C_{s}[f]$ is estimated as $O(\partial)$ or $O(\Omega)$. While for $C[f]$, there is no such a suppression factor. 
 
Hereafter, we split $\mathcal{W}$ into the unpolarized part and polarized one. Note that if we  neglect the term linearized to spin $\bm{s}$ in Eq.(\ref{hsr}), the  transition rate turns out the familiar unpolarized form,
\begin{eqnarray}
\bar{\mathcal{W}}&\equiv& \delta^{(4)}(p+p^\prime-p_1-p_2)\,
\frac{1}{8} \sum_{r,r',r_1,r_2} \langle{p,p^\prime;r,r^\prime|t|p_1,p_2;r_1,r_2}\rangle
\langle{p_1,p_2;r_1,r_2|t^\dagger|p,p^\prime;r,r^\prime}\rangle \;.
\label{local_col_GLW_after2}
\end{eqnarray}
 Eq.(\ref{local_col_GLW_after2}) is nothing but our widely used local collision term of two-body scattering.
 
As a reminder, we comment that the spatial shift $\Delta$
captures the crucial nonlocality of collisions,
where $\hat{t}^\mu$ is the time-like unit vector which is $(1,\boldsymbol{0})$ 
in the frame where $p^\mu$ is measured.
To note, such an appealing structure of the collision kernel  originates from the nontrivial tensor structure of particle fields, or equivalently the nontrivial dynamics introduced by spin degrees of freedom compared to related discussions about the scalar field in \cite{Hu:2021plu}. 


  It is widely known that  it is hard to find a physical decomposition for total angular  momentum tensor valid for all possible systems and we can only choose one specific pseudo gauge appropriate for the physical system of our interests \cite{Hehl:1976vr}. When talking about dynamic spin polarization, it is the mechanism of spin-orbit angular momentum conversion that works. Consequently, we expect the spin angular momentum itself (as internal degree of freedom) is conserved when interactions are turned off. While it is not when interactions are turned on. The pseudo-gauge proposed by Hilgevoord and Wouthuysen (HW) exactly satisfies the requirement \cite{Speranza:2020ilk}. Thus it is reasonable and natural to take HW gauge for our investigating spin polarization in the present work. With our choice, 
  the relevant  particle current, energy-momentum tensor and spin tensor are defined as
 \begin{align}
 \label{N}
 &N^{\mu}\equiv\int\,d\Gamma \,p^\mu f(x,p,\bm{s}),\\
 \label{T}
 &T_{\text{HW}}^{\mu\nu}\equiv\int \,d\Gamma \,p^\mu p^\nu f(x,p,\bm{s}),\\
 \label{S}
 &S_{\text{HW}}^{\lambda,\mu\nu}\equiv\int \,d\Gamma \,p^\lambda (\frac{1}{2}\Sigma^{\mu\nu}_{\bm{s}}-\frac{1}{2m^2}p^{[\mu}\partial^{\nu]} )f(x,p,\bm{s}),
 \end{align}
 where the antisymmetric part of Eq.(\ref{T}) is omitted because it is at the second order of gradients $O(\partial^2)$ \cite{Speranza:2020ilk} and outside the range of our following linear order analysis. 
 In the following sections, when nothing confusing occurs, the subscript HW will be omitted.
 
The above tensors can be conveniently decomposed as,
 \begin{align}
 \label{N1}
 &N^{\mu}=nu^\mu+V^\mu,\\
 \label{T1}
 &T^{\mu\nu}=eu^\mu u^\nu-P\Delta^{\mu\nu}+\pi^{\mu\nu}+\Pi\Delta^{\mu\nu},\\
\label{S1}
  &S^{\lambda,\mu\nu}=u^\lambda S^{\mu\nu}+\delta S^{\lambda,\mu\nu},
 \end{align}
 with $n$ the particle number density, $e$ the energy density, $P$ the static pressure and $S^{\mu\nu}$ the spin density. The dissipative quantities $V^\mu, \pi^{\mu\nu}$ and $\Pi$ are the diffusion current, shear stress tensor, bulk viscous pressure respectively and $V^\mu u_\mu=\pi^{\mu\nu}u_\mu=0$. Note that we have chosen Landau choice of fluid velocity and  imposed Landau matching conditions by requiring 
   \begin{align}
   T^{\mu\nu}u_\nu=eu^\mu, \quad u_\mu N^\mu=u_\mu N^\mu_{\text{eq}}, \quad u_\mu T^{\mu\nu}u_\nu=u_\mu T^{\mu\nu}_{\text{eq}}u_\nu.
   \end{align}
  Also another Landau matching condition for unambiguously defining $S^{\mu\nu}$  is taken,
 \begin{align}
 \label{J}
 u_\lambda J^{\lambda,\mu\nu}=u_\lambda J_{\text{eq}}^{\lambda,\mu\nu},
 \end{align}
 where the conserved tensor $J^{\lambda,\mu\nu}\equiv S^{\lambda,\mu\nu}+x^\mu T^{\lambda\nu}-x^\nu T^{\mu\lambda}$ is total angular momentum tensor. In the present case, HW 
 energy-momentum tensor takes only symmetric form, the matching condition Eq.(\ref{J}) degenerates into 
  \begin{align}
  \label{S12}
  u_\lambda S^{\lambda,\mu\nu}=u_\lambda S_{\text{eq}}^{\lambda,\mu\nu}.
  \end{align}

 \section{Linear collision operator}
 \label{linear}	
 
It is shown in \cite{Weickgenannt:2021cuo, Weickgenannt:2020aaf,Hu:2022xjn} that the collision term Eq.(\ref{cf}) is consistent with  local equilibrium distribution function \cite{Becattini:2013fla,Florkowski:2017ruc},
\begin{align}
\label{feq}
&f_{\text{leq}}(x,p,\bm{s})=\frac{1}{(2\pi)^3}\exp[\xi-\beta\cdot p+\frac{\Omega_{\mu\nu}\Sigma_{\bm{s}}^{\mu\nu}}{4}], 
\end{align}
where $\Omega_{\mu\nu}$ denotes spin potential,  the dipole moment $\Sigma_{\bm{s}}^{\mu\nu}$ is defined as  $\Sigma_{\bm{s}}^{\mu\nu}\equiv -\frac1m \epsilon^{\mu\nu\alpha\beta}p_\alpha \bm{s}_\beta$, and  $\beta^\mu\equiv\frac{u^\mu}{T},\xi\equiv \frac{\mu}{T}$ with the temperature $T$,  and  the chemical potential $\mu$. Without details, we summarize the conditions for global equilibrium distribution as
\begin{align}
\label{condition}
&\beta^\mu=a^\mu+\Omega^{\mu\nu}x_\nu,\quad a^\mu=\text{const}.
\end{align}
 Here we want to emphasize that such a conclusion relies on the assumption of conservative total angular momentum $J^{\mu\nu}$,
\begin{align}
J^{\mu\nu}=2\Delta^{[\mu}p^{\nu]}+\frac{1}{2}\Sigma^{\mu\nu}_{\bm{s}},
\end{align}
because the collision term itself does not conserve  total angular  momentum for lack of  a similar delta function like $ \delta^{(4)}(p+p^\prime-p_1-p_2)$. From now on, we follow this necessary assumption, and some modification needs to be built into Eqs.(\ref{local_col_GLW_after}) and (\ref{local_col_GLW_after2}) based on phenomenological consideration. Then Eq.(\ref{local_col_GLW_after2}) is substituted by
\begin{eqnarray}
\bar{\mathcal{W}}\times F[p,p^\prime,p_1,p_2;\bm{s},\bm{s}^\prime,\bm{s}_1,\bm{s}_2],
\label{local_col_GLW_after3}
\end{eqnarray}
where the dimensionless function $F$ is inserted respecting the conservation for the total angular momentum and we leave the detailed form of $F$ unspecified. Aside from this constraint on $F$, we require that $F$ respect the symmetry of interchanging the variables of initial states with that of final states, which is not a severe constraint. One can naively treat $F$ as a function proportional to  $\delta^{(6)}(J^{\mu\nu}+J^{\prime\mu\nu}-J_1^{\mu\nu}-J_2^{\mu\nu})$ with the superscript reminding us of six degrees of freedom of $J^{\mu\nu}$. In addition, such an insertion needs also to be made into Eq.(\ref{local_col_GLW_after}). In  Appendix.\ref{prove}, we prove that the linearized form of the collision kernel is Hermitian and non-negative with the unpolarized transition rate like Eq.(\ref{local_col_GLW_after3}). There are  two points that deserve special attention. One is that the unpolarized transition rate respects the principle of detailed balance. The other is that the non-negative property is important because it selects kinetic modes  protected by conservation laws, which is reported in details in Sec.\ref{relax}. Therefore, we concentrate only on the non-negative Hermitian linearized collision operator afterwards.

 After discussing the equilibrium state,  we can  linearize the transport equation Eq.(\ref{boltz}) around the global equilibrium state $f_G$. Before doing that, Eq.(\ref{boltz}) can be explicitly decomposed into
\begin{align}
\label{boltz1}
&p\cdot u Df(x,p,\bm{s})+ p^\nu\nabla_\nu f(x,p,\bm{s})=C[f],
\end{align}
and the distribution function is 
\begin{align}
\label{fk}
&f\equiv f_G(1+\chi(x,p,\bm{s})),  
\end{align}
where $f_G$ is exactly the global equilibrium distribution, equivalently,  Eq.(\ref{feq}) satisfying Eq.(\ref{condition}).
Insert  Eq.(\ref{fk}) into Eq.(\ref{boltz1}), and we get
\begin{align}
\label{cf1}							
&p^\mu u_\mu D\chi(x,p,\bm{s})+p^\mu\nabla_\mu \chi(x,p,\bm{s})=-\mathcal{L}[\chi], 
\end{align}
with the linearized collision operator defined  as:
\begin{eqnarray}
\mathcal{L}[\phi]
&\equiv &\frac{1}{(2\pi)^3}(1+\frac{\Omega_{\mu\nu}\Sigma_{\bm{s}}^{\mu\nu}}{4})^{-1} \int d\Gamma^\prime d\Gamma_1 d\Gamma_2  \,
\bar{\mathcal{W}} \, F[p,p^\prime,p_1,p_2;\bm{s},\bm{s}^\prime,\bm{s}_1,\bm{s}_2] \exp(\xi-a\cdot p^\prime\,)\nn\\
& &\times  \Big[\frac 1 2\Omega_{\mu\nu}(J^{\mu\nu}+J^{\prime\mu\nu})\,\big(\phi(x,p,\bm{s})+\phi(x,p^\prime,\bm{s}^\prime)-\phi(x,p_1,\bm{s}_1)-\phi(x,p_2,\bm{s}_2)\,\big)\nn\\
&&+\left(\phi(x+\Delta,p,\bm{s})+\phi(x+\Delta^\prime,p^\prime,\bm{s}^\prime) -\phi(x+\Delta_1,p_1,\bm{s}_1) -\phi(x+\Delta_2,p_2,\bm{s}_2)  \right)\Big].
\label{colleqq1}
\end{eqnarray}
For simplicity, the background profile is taken to be $u^{\mu}=(1,0,0,0), \Omega^{\mu\nu}=0$.
Here we note as an aside that the background profile can absolutely  be chosen to be various. For example, a finite vorticity can survive in global equilibrium, then $\beta$ field can possess coordination dependence. In the meantime, thermodynamic integrals determining the dispersion relations shown in the Appendix are totally different and more complicated because  $\hat{t}$ does not equal to $u$ any more. On the other hand, only spin exchange collision term should be taken into account when $\Omega$ is finite.  In order to compare with similar analysis on hydrodynamic modes in \cite{Hattori:2019lfp}, the calibration should be made ensuring the global equilibrium configurations are the same, i.e., $u^{\mu}=(1,0,0,0), \Omega^{\mu\nu}=0$. Thus $a^\mu=\beta^\mu=\beta u^\mu$ with $\beta\equiv \frac{1}{T}$ and the linearized collision operator is greatly simplified. 


 Considering that $\phi(x+\Delta)$  is actually the linear approximation to $\phi(x)+\Delta\cdot \partial\phi$ \cite{Weickgenannt:2021cuo},
so we return to its initial form and naturally split the linearized collision term into a local one
\begin{eqnarray}
\mathcal{L}_1[\phi]
&\equiv &\frac{1}{(2\pi)^3} \int d\Gamma^\prime d\Gamma_1 d\Gamma_2  \,
\bar{\mathcal{W}} \, F[p,p^\prime,p_1,p_2;\bm{s},\bm{s}^\prime,\bm{s}_1,\bm{s}_2]
 \exp(\xi-\beta\cdot p^\prime\,)\nn\\ &\times&\Big[\phi(x,p,\bm{s})+\phi(x,p^\prime,\bm{s}^\prime)-\phi(x,p_1,\bm{s}_1)-\phi(x,p_2,\bm{s}_2)\,\Big],
\label{colleqqcl}
\end{eqnarray}
and a nonlocal one featured by the nonlocal shift $\Delta$
  \begin{eqnarray}
  \mathcal{L}_2[\phi]
  &\equiv &\frac{1}{(2\pi)^3} \int d\Gamma^\prime d\Gamma_1 d\Gamma_2  \,
  \bar{\mathcal{W}}  F[p,p^\prime,p_1,p_2;\bm{s},\bm{s}^\prime,\bm{s}_1,\bm{s}_2]\,  \exp(\xi-\beta\cdot p^\prime\,)\nn\\
  & \times & \Big[\Delta\cdot\partial_x\phi(x,p,\bm{s})+\Delta^\prime\cdot\partial_x\phi(x,p^\prime,\bm{s}^\prime) -\Delta_1\cdot\partial_x\phi(x,p_1,\bm{s}_1) -\Delta_2\cdot\partial_x\phi(x,p_2,\bm{s}_2)  \Big].
  \label{colleqqs1}
  \end{eqnarray}
 
As is exhibited in Appendix.(\ref{prove}), $\mathcal{L}_1$ is an Hermitian operator while $\mathcal{L}_2$ is not for the non-commutative property of the derivative.
With our proposed division, Eq.(\ref{cf1}) is transformed into
 \begin{align}
 \label{cf3}
 &p^\mu u_\mu D\chi(x,p,\bm{s})+p^\mu\nabla_\mu \chi(x,p,\bm{s})+\mathcal{L}_2[\chi]=-\mathcal{L}_1[\chi].
 \end{align}
 
It is convenient to introduce the Fourier transformation of $\chi(x,p,\bm{s})$
\begin{align}
\tilde{\chi}(k,p,\bm{s})=\int d^4 x\exp(ik\cdot x)\chi(x,p,\bm{s}),
\end{align}
and cast Eq.(\ref{cf1}) into
\begin{align}
\label{cf2}
&\tau \omega\tilde{\chi}+\hat{p}^\mu \kappa_\mu \tilde{\chi}+L_2[\tilde{\chi}]=-iL_1[\tilde{\chi}], 
\end{align}
where the following notations are utilized 
\begin{align}
\label{pert}
\tau\equiv \frac{p\cdot u}{T},\quad \omega\equiv \frac{u\cdot k}{n\sigma(T)},\quad\hat{p}\equiv\frac{p}{T},\quad \kappa^\alpha\equiv\frac{\Delta^{\alpha\beta}k_\beta}{n\sigma(T)},\quad \kappa\equiv\sqrt{-\kappa\cdot\kappa},\quad l^\alpha\equiv\frac{\kappa^\alpha}{\kappa},
\end{align}
and  $\sigma(T)$ is an arbitrary constant with the dimension of  total cross sections.
 The dimensionless collision operator is,
\begin{eqnarray}\label{colleqqcl1}
L_1[\phi]
&\equiv &\frac{\exp(\xi)}{(2\pi)^3n\sigma(T)T} \int d\Gamma^\prime d\Gamma_1 d\Gamma_2  \,
\bar{\mathcal{W}} \,  F[p,p^\prime,p_1,p_2;\bm{s},\bm{s}^\prime,\bm{s}_1,\bm{s}_2] \exp(-\beta\cdot p^\prime\,)\nn\\
&\times & \Big[\phi(k,p,\bm{s})+\phi(k,p^\prime,\bm{s}^\prime)-\phi(k,p_1,\bm{s}_1)-\phi(k,p_2,\bm{s}_2)\,\Big],\nn\\
\end{eqnarray}
and at the meanwhile,
  \begin{eqnarray}
L_2[\phi]
  &\equiv &\frac{\exp(\xi)}{(2\pi)^3T} \int d\Gamma^\prime d\Gamma_1 d\Gamma_2  \,
  \bar{\mathcal{W}} \,   F[p,p^\prime,p_1,p_2;\bm{s},\bm{s}^\prime,\bm{s}_1,\bm{s}_2]\exp(-\beta\cdot p^\prime\,)\nn\\
  & \times&  \Big[\Delta\cdot \kappa\phi(k,p,\bm{s})+\Delta^\prime\cdot \kappa\phi(k,p^\prime,\bm{s}^\prime) -\Delta_1\cdot  \kappa\phi(k,p_1,\bm{s}_1) -\Delta_2\cdot \kappa \phi(k,p_2,\bm{s}_2)  \Big].
  \label{colleqqs2}
  \end{eqnarray}
  Here the inner product is  global equilibrium state given by
  \begin{align}
  \label{inner}
  (B,C)=\frac{1}{(2\pi)^3}\int d\Gamma \exp(\xi-\beta\cdot p)B(p,\bm{s}) C(p,\bm{s}).
  \end{align}
 To proceed, the adjoint operator of $L_2$ is denoted by $L^\dagger_2$ and defined such that
  \begin{align}
  \label{L2}
  (L^\dagger_2[B],C)=(B,L_2[C]).
  \end{align}
  By utilizing the Hermitian property of $L_1$, one can show that
  \begin{align}
  (B,L_2[C])=n\sigma(T)(B,L_1[\Delta\cdot\kappa C])=n\sigma(T)(L_1[B],\Delta\cdot\kappa C)=n\sigma(T)(\Delta\cdot\kappa L_1[B],C).
  \end{align}
  Therefore, the adjoint operator $L^\dagger_2$ is identified with $n\sigma(T)\Delta\cdot\kappa L_1$ and explicitly takes the following form
  \begin{align}
  L^\dagger_2[\phi]
  &\equiv \frac{\exp(\xi)}{(2\pi)^3T} \int d\Gamma^\prime d\Gamma_1 d\Gamma_2  \,
  \bar{\mathcal{W}} \, F[p,p^\prime,p_1,p_2;\bm{s},\bm{s}^\prime,\bm{s}_1,\bm{s}_2] \exp(-\beta\cdot p^\prime\,)\nn\\
  &\times\Delta\cdot\kappa \Big[\phi(k,p,\bm{s})+\phi(k,p^\prime,\bm{s}^\prime)-\phi(k,p_1,\bm{s}_1)-\phi(k,p_2,\bm{s}_2)\,\Big],
  \end{align}
  with the same argument in Appendix.\ref{prove}, $L_2$ is also Hermitian.
   Although $\mathcal{L}_2$ is not Hermitian,  $L_2$ turns out to be an Hermitian operator for the replacement of the derivative in Eq.(\ref{colleqqs1}) by $\kappa$ in Eq.(\ref{colleqqs2}).

   \section{Degenerate perturbation theory and linear mode analysis}
   \label{expansion}
   As is stressed previously, solving normal modes can be treated in the fashion as used in quantum mechanics. Here we regard the $p\cdot \kappa$ and $L_2$ terms in Eq.(\ref{cf2})  as a perturbation with respect to the linearized collision operator $-iL_1$, equivalently, we take the non-uniformity $\kappa$ as  a small quantity.
  Then the unperturbed equation reduces to an eigenvalue problem,
\begin{align}
\label{unpert}
&-iL_1\tilde{\chi}^{(0)}= \tau\omega^{(0)}\tilde{\chi}^{(0)}.
\end{align}
 It is easy to show that $11$ zero modes (or collision invariants) $1, p^\mu$ and $J^{\mu\nu}$ trivially solve the zeroth order equation.  Compared to ordinary hydrodynamic theory, the six new modes  are related to the conservation of  total angular momentum in nonlocal collisions between particles.  Noting we  have exhausted the global symmetries relevant for present research, there should be no more conserved currents and no more zero modes.

When it comes to  the first-order perturbation of $p\cdot \kappa$ and $L_2$, we denote the $n$-th order contribution with a superscript $n$ and only focus on  eleven zero  modes which are in tight connection with fluid-dynamic variables. 
The eigenfunctions set  for eleven-fold degenerate zeros  can be readily chosen as
\begin{align}
\label{sets}
&\tilde{\psi}_{1}=\frac{1}{\sqrt{V_{1,1}}}, \quad \tilde{\psi}_{2}=\beta\frac{u\cdot p-\frac{e}{n}}{\sqrt{V_{2,2}}},\quad \tilde{\psi}_{3}=\frac{\beta l\cdot p}{\sqrt{V_{3,3}}},\quad \tilde{\psi}_{4}=\frac{\beta j\cdot p}{\sqrt{V_{3,3}}}, \quad \tilde{\psi}_{5}=\frac{\beta v\cdot p}{\sqrt{V_{3,3}}},\nn\\
&\tilde{\psi}_{6}=\frac{u_\mu J^{\mu\nu}l_\nu}{\sqrt{V_{6,6}}},\quad \tilde{\psi}_{7}=\frac{u_\mu J^{\mu\nu}j_\nu}{\sqrt{V_{6,6}}},\quad \tilde{\psi}_{8}=\frac{u_\mu J^{\mu\nu}v_\nu}{\sqrt{V_{6,6}}},\quad \tilde{\psi}_{9}=\frac{l_\mu J^{\mu\nu}j_\nu}{\sqrt{V_{9,9}}},\nn\\
&\tilde{\psi}_{10}=\frac{l_\mu J^{\mu\nu}v_\nu}{\sqrt{V_{9,9}}}, \quad \tilde{\psi}_{11}=\frac{j_\mu J^{\mu\nu}v_\nu}{\sqrt{V_{9,9}}},
\end{align}
where Schmidt orthogonalization  is used
and these eigenfunctions satisfy the orthonormal condition
\begin{align}
\label{one}
(\tilde{\psi}_\alpha,\tau \tilde{\psi}_\beta)=\delta_{\alpha\beta}.
\end{align}
The definitions of two auxiliary vectors $j, v$ and the normalized factor $V_{i,j}$ are all defined in Appendix.\ref{int1}.

Taking into account the first-order perturbation, we obtain the inhomogeneous integral equation for $\tilde{\chi}^{(1)}$ 
\begin{align}
\label{pert1}
&-iL_1\tilde{\chi}_\alpha^{(1)}= \tau\omega_\alpha^{(1)}\tilde{\chi}_\alpha^{(0)}+\hat{p}^\mu \kappa_\mu \tilde{\chi}_\alpha^{(0)},  
\end{align}
where the property of $L_2\tilde{\chi}_\alpha^{(0)}=0$ is used.
According to  degenerate perturbation theory, the solubility condition is 
\begin{align}
\label{soluble}
(\tilde{\psi}_\alpha,\,\tau\omega^{(1)}\tilde{\chi}^{(0)}+p^\mu \kappa_\mu \tilde{\chi}^{(0)})=0,
\end{align}
where $\tilde{\chi}^{(0)}$ should be understood as the linear combination of the eigenfunctions $\tilde{\psi}_\beta$ in Eq.(\ref{sets}).
 Therefore the frequency $\omega$  obeys the dispersion relation, i.e., the secular equation,
\begin{align}
\label{secu}
& 
\left|\begin{array}{cccccccccccc}
\omega^{(1)}&0&H_{1,3}&0&0&0  & 0 & 0&0&0&0\\
 0&\omega^{(1)}&H_{2,3}&0&0&0 & 0& 0&0&0&0\\
 H_{1,3}&H_{2,3}&\omega^{(1)}&0&0&0& 0 &  0&0&0&0\\
 0&0&0&\omega^{(1)}&0&0  &0 & 0  &0&0&0\\
 0&0&0&0&\omega^{(1)}&0  &0 & 0  &0&0&0\\
 0&0&0&0&0&\omega^{(1)}  & 0 & 0&0&0&0\\
 0&0&0&0&0&0 & \omega^{(1)}& 0&H_{7,9}&0&0\\
 0&0&0&0&0&0& 0 &  \omega^{(1)}&0&H_{7,9}&0\\
 0&0&0&0&0&0  &H_{7,9} & 0  &\omega^{(1)}&0&0\\
 0&0&0&0&0&0  &0 & H_{7,9}  &0&\omega^{(1)}&0\\
 0&0&0&0&0&0  &0 & 0  &0&0&\omega^{(1)}\\
\end{array}\right|=0 \,,
\end{align}
where matrix elements $H_{i,j}$ are given in Appendix.\ref{pertu}.

The solution to  this equation is made up of the eigenvalues
\begin{align}
&\omega^{(1)}_1=-\omega^{(1)}_2=\sqrt{H^2_{2,3}+H^2_{1,3}}{},\quad \omega^{(1)}_{3}=\omega^{(1)}_{4}=\omega^{(1)}_{5}=0,\nn\\
&\omega^{(1)}_6=\omega^{(1)}_{11}=0,\quad\omega^{(1)}_7=\omega^{(1)}_8=H_{7,9},\quad \omega^{(1)}_{9}=\omega^{(1)}_{10}=-H_{7,9},
\end{align}
and the eigenfunctions
\begin{align}
\label{chi}
&\tilde{\chi}^{(0)}_1=\frac{1}{\sqrt{2}}(-\frac{H_{1,3}}{\sqrt{H^2_{1,3}+H^2_{2,3}}}\tilde{\psi}_1-\frac{H_{2,3}}{\sqrt{H^2_{1,3}+H^2_{2,3}}}\tilde{\psi}_2+\tilde{\psi}_3),\nn\\
&\tilde{\chi}^{(0)}_2=\frac{1}{\sqrt{2}}(\frac{H_{1,3}}{\sqrt{H^2_{1,3}+H^2_{2,3}}}\tilde{\psi}_1+\frac{H_{2,3}}{\sqrt{H^2_{1,3}+H^2_{2,3}}}\tilde{\psi}_2+\tilde{\psi}_3),\nn\\
&\tilde{\chi}^{(0)}_3=\frac{1}{\sqrt{\frac{H_{2,3}^2}{H^2_{1,3}}+1}}(-\frac{H_{2,3}}{H_{1,3}}\tilde{\psi}_1+\tilde{\psi}_2),\quad \tilde{\chi}^{(0)}_4=\tilde{\psi}_4,\quad \tilde{\chi}^{(0)}_5=\tilde{\psi}_5,\quad \tilde{\chi}^{(0)}_6=\tilde{\psi}_6,\quad\nn\\
& 
\tilde{\chi}^{(0)}_{7}=\frac{1}{\sqrt{2}}(\tilde{\psi}_7-\tilde{\psi}_9),\quad  \tilde{\chi}^{(0)}_{8}=\frac{1}{\sqrt{2}}(\tilde{\psi}_8-\tilde{\psi}_{10}),\quad\tilde{\chi}^{(0)}_{9}=\frac{1}{\sqrt{2}}(\tilde{\psi}_7+\tilde{\psi}_9),\quad  \tilde{\chi}^{(0)}_{10}=\frac{1}{\sqrt{2}}(\tilde{\psi}_8+\tilde{\psi}_{10}),\quad \tilde{\chi}^{(0)}_{11}=\tilde{\psi}_{11}.
\end{align}
It is not hard to verify that the results of the first five modes are exactly those in \cite{DeGroot:1980dk}. As a consistent check, we find the eigen functions $\tilde{\chi}^{(0)}_\alpha,\,\alpha=6\cdots,11$ indeed satisfy the solubility condition Eq.(\ref{soluble}) with $\omega_\alpha^{(1)}=0$.
 
 Likewise, we obtain again the  solubility condition for second-order perturbation equation,
\begin{align}
\label{soluble2}
(\tilde{\chi}^{(0)}_\alpha,\,\tau\omega_\beta^{(1)}\tilde{\chi}^{(1)}_\beta+\hat{p}^\mu \kappa_\mu \tilde{\chi}^{(1)}_\beta+\tau\omega_\beta^{(2)}\tilde{\chi}_\beta^{(0)})=0,
\end{align}
where we have used Eq.(\ref{L2}) to vanish the inner product of $\tilde{\chi}^{(0)}_\alpha$ and $L_2\tilde{\chi}_\beta^{(1)}$. In order to form a comparison with related results in \cite{DeGroot:1980dk}, the second order frequencies can be written as with the assistance of the bracket notation,
\begin{align}
\label{bracket}
\omega^{(2)}_\alpha=i[\tilde{\chi}^{(1)}_\alpha,\tilde{\chi}^{(1)}_\alpha],
\end{align}
where $\tilde{\chi}^{(1)}$ should be obtained by solving the integral equation Eq.(\ref{pert1})  and the bracket  notation is given by $[B,C]\equiv (L_1[B],C)$.
The second-order frequencies in the form of the bracket notation is in accordance  with \cite{DeGroot:1980dk}. 

 
As is seen from the secular equation Eq.(\ref{secu}), the spin modes and spin-less modes decouple from each other, thus the dispersion relations for the spin-less modes receive no corrections from the spin effects and retain the same form as that in the linear analysis of the ordinary hydrodynamics, which is also confirmed in \cite{Hattori:2019lfp}. Therefore, the frequencies of the spinless modes up to the second-order in $\kappa$ is already available as follows \cite{DeGroot:1980dk}
\begin{align}
&\omega_1=\sqrt{\frac{\gamma}{\hat{h}}}\kappa-i\frac{n\sigma}{2h}\Big(\frac{4}{3}\eta+\zeta+\frac{\big((\gamma-1)\hat{h}-\gamma\,\big)^2}{\gamma\hat{h}}\lambda\Big)\kappa^2,\nn\\
&\omega_2=-\sqrt{\frac{\gamma}{\hat{h}}}\kappa-i\frac{n\sigma}{2h}\Big(\frac{4}{3}\eta+\zeta+\frac{\big((\gamma-1)\hat{h}-\gamma\,\big)^2}{\gamma\hat{h}}\lambda\Big)\kappa^2,\nn\\
& \omega_{3}=-i\frac{\gamma-1}{\gamma}\lambda\sigma\kappa^2,\quad \omega_{4}=\omega_{5}=-i\frac{n\sigma}{h}\eta\kappa^2,
\end{align}
with the transport coefficients shear viscosity, bulk viscosity, and diffusion coefficient denoted by $\eta,\,\zeta$ and $\lambda$ respectively, $\hat{h}$  the reduced enthalpy density $\hat{h}\equiv\frac{e+P}{T}$, and $\gamma$ denoting the ratio of the heat capacities at constant pressure $c_p\equiv\big(\frac{\partial (h/n)}{\partial T}\big)_p$ and at constant volume $c_v\equiv\big(\frac{\partial (e/n)}{\partial T}\big)_v$. As a low-energy effective theory, the kinetic theory should match  the coarse-grained hydrodynamics in the limit of long wavelengths. The one-to-one correspondence between hydrodynamic modes and kinetic modes has a great practical importance as the basis for a theory of transport coefficients, which should be thought of  from a more ``philosophical" view.  A fluid system is a collection of particles moving in a quite disorder manner, however in the limit of long wavelengths, the only possible modes of motion of the fluid are ordered modes such as  a sound-wave propagation, which originate from the dominant effect of the collisions \cite{Balescu, Resi}.

 In a short summary, among the five spinless hydrodynamic  modes there are two sound modes traveling with opposite sound speed and  the same damping rate, one purely decaying heat mode and two degenerate shear modes. 
On the other hand, the frequencies for the other six spin-related modes are
\begin{align}
&\omega_6=i[\tilde{\chi}^{(1)}_6,\tilde{\chi}^{(1)}_6],\quad\omega_{11}=i[\tilde{\chi}^{(1)}_{11},\tilde{\chi}^{(1)}_{11}],\quad\quad\omega_8=\omega_7=H_{7,9}+i[\tilde{\chi}^{(1)}_7,\tilde{\chi}^{(1)}_7],\quad \omega_{10}=\omega_{9}=-H_{7,9}+i[\tilde{\chi}^{(1)}_9,\tilde{\chi}^{(1)}_9].
\end{align}

By a careful comparison, we find  that these spin-related dispersion relations cannot match with those in \cite{Hattori:2019lfp}. The reason is that in that work the authors concentrates on the non-conservative spin density while our spin modes are inherently protected by conservation laws. There are four  propagating transverse spin modes, two degenerate modes with  $H_{7,9}/\kappa$  as the propagating speed damp at the rate  $-[\tilde{\chi}^{(1)}_7,\tilde{\chi}^{(1)}_7]$  while the other two propagate in the opposite direction with a damping rate of $-[\tilde{\chi}^{(1)}_9,\tilde{\chi}^{(1)}_9]$. On the other hand there are no propagating spin modes in \cite{Hattori:2019lfp}. In addition, the longitudinal modes are purely decaying at their respective damping rates. However, the bracket notation is rather abstract and not directly related to our familiar physical quantities or formula like spinless modes. To put it less abstract, first we notice that 
 \begin{align}
 \label{bzd}
 \omega^{(2)}_\alpha=-i\big( \tilde{\chi}^{(0)}_\alpha (\hat{p}^\mu \kappa_\mu+\tau\omega^{(1)}_\alpha ),\, L^{-1}_1\big[(\hat{p}^\mu \kappa_\mu+\tau\omega^{(1)}_\alpha )\tilde{\chi}^{(0)}_\alpha\big]\,\big).
 \end{align}
 If neglecting the non-uniformity (this term is an order of magnitude smaller), Eq.(\ref{cf3}) takes the form of the time evolution,
 \begin{align}
 \label{evolu}
  &\beta\tau D\chi(t,p,\bm{s})=-L_1[\chi],
 \end{align}
 when choosing the rest configuration $D=\partial t$ without losing the generality,
 the solution to the equation of evolution is shown as
 \begin{align}
 \label{evo}
\chi(t,p,\bm{s})=\exp(-\frac{L_1}{\beta\tau}t)\chi(p,\bm{s}),
 \end{align}
 with $\chi(t=0,p,\bm{s})=\chi(p,\bm{s})$.
 Therefore, Eq.(\ref{bzd}) can be cast into
  \begin{align}
  \label{bzd1}
  \omega^{(2)}_\alpha&=-i\big( \tilde{\chi}^{(0)}_\alpha (\hat{p}^\mu \kappa_\mu+\tau\omega^{(1)}_\alpha ),\, L^{-1}_1\big[(\hat{p}^\mu \kappa_\mu+\tau\omega^{(1)}_\alpha )\tilde{\chi}^{(0)}_\alpha\big]\,\big)\nn\\
  &=-i\kappa^2\big( \tilde{\chi}^{(0)}_\alpha (\hat{p}^\mu l_\mu+\tau\frac{\omega^{(1)}_\alpha}{\kappa} ) ,\,L^{-1}_1\big[(\hat{p}^\mu l_\mu+\tau\frac{\omega^{(1)}_\alpha}{\kappa} )\tilde{\chi}^{(0)}_\alpha\big]\,\big)\nn\\
    &=i\frac{1}{\beta(2\pi)^3}\int d\Gamma dt \exp(\xi-\beta\cdot p)\frac{1}{\tau} g(0,p,\bm{s})\exp(-\frac{L_1}{\beta\tau}t)g(0,p,\bm{s})\nn\\
  &=i\frac{1}{\beta(2\pi)^3}\int d\Gamma dt \exp(\xi-\beta\cdot p)\frac{1}{\tau} g(0,p,\bm{s})g(t,p,\bm{s})=i\frac{1}{\beta}\int_0^{\infty} dt\langle g(0,p,\bm{s})g(t,p,\bm{s})\rangle_0
  \end{align}
  with the definition $g(t=0,p,\bm{s})=(\hat{p}^\mu l_\mu+\tau\frac{\omega^{(1)}_\alpha}{\kappa} )\tilde{\chi}^{(0)}_\alpha$ and the expectation value in equilibrium identified as $\langle A\rangle_0\equiv\frac{1}{(2\pi)^3}\int d\Gamma \exp(\xi-\beta\cdot p)\frac{1}{\tau}A$. 
  
  From the Green-Kubo like formula derived here, we can find that the second-order frequencies can all be related to the time correlation functions. It is widely believed that the transport coefficients can be expressed by the Green-Kubo formula, based on which the second-order dispersion relations of spin modes are supposed to be used as definitions for new transport coefficients or their combination. Since there is no well-calibrated spin hydrodynamic theory  relevant to the spin kinetic theory adopted here and no corresponding definitions for our proposed novel  transport coefficients elsewhere, we have to content ourselves with ending with Eq.(\ref{bzd1}).

 \section{The relaxation of spin}
 \label{relax}
 In this section, we show how to relate our linear analysis with the relaxation of spin towards the equilibrium state. To that end, first we note that the imaginary parts of spinless hydrodynamic modes encode the information of the relaxation of conserved currents, e.g, the energy-momentum tensor and the particle number current marked by the hydrodynamic transport coefficients. Similarly, there is no doubt that this conclusion  can be generalized to new spin modes, encoding related information of the relaxation of total angular momentum.
 In the community of heavy-ion collisions, the relaxation of spin density towards equilibrium is important and interesting stimulated by the experimental research on local spin polarization. Whether the spin density relaxes to their equilibrium value earlier than other hydrodynamic variables or not is under debate. With the spin equilibrated  picture still undetermined, further numerical simulations or related modelings lack the stable basis. Inspired by the enlightening discussions about the ordinary  hydrodynamic modes analysis, we start with Eq.(\ref{evo}), then a general fluctuation i.e., the deviation function from the equilibrium
 distribution can be expanded with the orthogonal normalized sets of the eigenfunctions of linearized collision (evolution) operator,
\begin{align}
\label{t1}
\chi(t,p,\bm{s})=\sum_{\alpha=1}^{\infty}\psi_\alpha(  \psi_\alpha ,\,\chi)\approx\sum_{\alpha=1}^{11}\psi_\alpha ( \psi_\alpha ,\,\chi),
\end{align}
where $\psi_\alpha$ is the inverse Fourier transformation of $\tilde{\psi}_\alpha$  (they are actually the same because the thermodynamic parameters appearing in Eq.(\ref{sets}) are chosen to be independent of the coordination $x$ in global equilibrium) and the inner product $( \psi_\alpha ,\chi)$ represents the corresponding fluctuation amplitude (for instance, if $\alpha$ is chosen to be $1$, this inner product denotes the fluctuation amplitude of the particle number density).
Here we truncate the summation to first eleven terms for the first eleven zero modes are exactly the slowest modes which are  protected by the conversation laws.  Due to the non-negative property of the linearized collision operator, other modes with positive eigenvalues are damped by the exponential factor as time evolves in view of  Eq.(\ref{evo}). 

  After singling out  relevant zero modes from infinite many positive modes, it is our task to see how these zero modes response to the perturbation of non-uniformity, which will give us dispersion laws as is exhibited in the previous section. To that end, we  move back  to momentum space and the eigenfunctions used to expand the fluctuation function $\tilde{\chi}$ turn into $\tilde{\chi}_\alpha$ in Eq.(\ref{chi}),
\begin{align}
\label{chiw}
\tilde{\chi}(k,p,\bm{s})\approx\sum_{\alpha=1}^{11}\tilde{\chi}_\alpha( \tilde{\chi}_\alpha ,\,\tilde{\chi}).
\end{align}

Analogous to the definition of spin tensor Eq.(\ref{S}), we define a new fluctuation for spin density, 
\begin{align}
\label{deltas}
\delta S^{\mu\nu}(x,p,\bm{s})=u_\lambda \int \,\frac{d\Gamma}{(2\pi)^3} e^{\xi-\beta\cdot p}\,p^\lambda \frac{1}{2}\Sigma^{\mu\nu}_{\bm{s}}\chi(x,p,\bm{s}),
\end{align}
where we have invoked that $u_\lambda \int \,d\Gamma \,p^\lambda \frac{1}{2}\Sigma^{\mu\nu}_{\bm{s}}f_{\text{eq}}$ is the spin density in equilibrium compatible with the related  definition in \cite{Hattori:2019lfp} and we neglect the derivative term belonging to the smaller order in gradients. By analogy with $\delta T^{\mu\nu}$ or $\delta N^\mu$, the natural interpretation for $(  \frac{u\cdot p}{2}\Sigma^{\mu\nu}_{\bm{s}} ,\,\chi)$ is the fluctuation amplitude for spin density.

Considering the overlap of $\Sigma^{\mu\nu}_{\bm{s}}$ with the spinless eigen modes ($\alpha=1,2,3,4,5$) vanishes for $	\int dS(p)\, \bm{s}^\mu  =  0$, only six spin-related fluctuation amplitudes $(\chi_\alpha ,\chi)$  are responsible for the relaxation of spin density,
\begin{align}
\label{chiw1}
( \frac{u\cdot p}{2}\Sigma^{\mu\nu}_{\bm{s}} ,\,\tilde{\chi})=
\sum_{\alpha=6}^{11}( \frac{u\cdot p}{2}\Sigma^{\mu\nu}_{\bm{s}},\,\tilde{\chi}_\alpha )( \tilde{\chi}_\alpha ,\,\tilde{\chi}),
\end{align}
and these independent six modes decay at different relaxation times. Therefore the relaxation time for spin density fluctuation should be identified as the largest one, namely, 
\begin{align}
\label{tau}
\tau_s=\max\{\frac{1}{|\omega_\alpha^{(2)}|},\alpha=6,\cdots 11\}
\end{align}
where  $|A|$ denotes  the amplitude of a complex $A$. Given specific interaction and proper form for constraint function $F$ in Eq.(\ref{colleqq1}), solve the complicated integral equation Eq.(\ref{pert1}) and the relaxation times for ordinary hydrodynamic dissipative quantities and spin density can be determined, constructing the related hierarchy for the different relaxation times and clarifying the relevant equilibrium picture.

Before ending this section, some comments about any possible pseudo-gauge dependencies are presented as follows. As is put forward in previous sections, the energy momentum tensor in distinct gauges take the same symmetric form as far as only first order gradients are in consideration. The dispersion relations of spinless modes do not change when taking distinct pseudo-gauges.  However, the variation of gauge will lead to the variation of definition of spin density and alter the  spin density fluctuation Eq.(\ref{deltas}), too. The variation should be of the form $\sim \int \,dS(p) \, \bm{s}\chi(x,p,\bm{s})$ with only spin-related structure  displayed,  which indicates that involved modes responsible for spin relaxation are still unchanged and Eq.(\ref{tau}) is formally identical. On the other hand,  the precise relaxation rates of these spin modes are also unchanged given a fixed background profile on top of which the linear analysis is carried out. In a summary, our results presented before are independent of the choice of pseudo-gauges.

Note that our formalism is based on neglecting the antisymmetric part of HW energy momentum, which manifests that spin angular momentum and orbit angular momentum are separately conserved. Though the present framework is rather restrictive,  it is reasonable as $\partial_\mu T^{[\mu\nu]}$ is counted as $O(\partial^3)$ in the motion equation and thus negligible. 
Moreover, the spin Boltzmann equation derived from Wigner formalism is limited to first order in semi-classical expansion (or gradients expansion) and does not encode the information of  the antisymmetric part of HW energy momentum tensor by construction \cite{Weickgenannt:2022jes}. Therefore  the antisymmetric part of HW energy momentum tensor can not be expressed by distribution function $f$ or collision kernel $C[f]$ presented here and  the discussion on it is beyond the application range of present framework.
We comment that it is a more straightforward way to formulate the retarded correlators from underlying field theoretical calculation  and then systematically look for non-hydrodynamic modes related to non-conserved spin density. 
 
 \section{Summary and outlook}
 \label{su}
 We provide a discussion about  normal modes of linearized collision operator based on the spin kinetic theory for massive fermions  \cite{Weickgenannt:2021cuo} in this paper. By insisting on Hermitian and non-negative properties of linearized operator, we neglect the non-diagonal part of the transition rate and solve the dispersion relations for normal modes according to the degenerate perturbation theory we frequently met in quantum mechanics.
 With the conservation of total angular momentum in a collision event  phenomenologically considered, we find eleven zero modes which are protected by  conserved laws and thus are  in relation with  spin hydrodynamic theory.
 
  Following the similar procedures that are used in quantum mechanics,  we obtain the dispersion relations of these modes, among which the results of five spinless modes are consistent with well-known conclusions. On the other hand, the frequencies of remaining spin-related modes are also solved up to second order in wave vector though the second-order corrections are only  formal solutions, which are proved to be in relation with the time correlation functions. In other words, one can directly calculate the time correlation function without recourse to solving the complicated integral equation. We also show that our framework can be applied to investigate the relaxation of  spin density. Our findings manifest that the relaxation of spin density has nothing to do with the spinless modes and the relaxation time is identified as the largest one of the reciprocals of damping rates for those spin-related modes. Therefore, given specific interactions and reasonable parameterized form for the constraint function in Eq.(\ref{colleqqcl1}), the relaxation for both spin density and other dissipative hydrodynamic quantities can be  determined, which at least provides a comparison for these typical time scales and decide which one is the slow process. The clarification of the hierarchy for relaxation times based on reliable quantum kinetic theory is highly non-trivial in resolving  the problem of discovering the local spin polarization in the experiments of relativistic heavy-ion collisions. Note that spin Boltzmann equation adopted here does not encode the antisymmetric part of HW energy momentum tensor, therefore spin is approximately conserved and spin-orbit conversion happens at the second order in gradients expansion. 
  
  There is also one thing that deserves attention. Because our discussion is limited to ignoring the polarized part in the transition rate, thus the linearized collision operator  is Hermitian. Without this approximation, the unperturbed operator is not Hermitian, the eigenvalues are not necessarily real while the eigenfunctions are not necessarily orthogonal, which means we can't solve the problem following the same fashion as used in quantum mechanics. For the impact of the polarized transition rate on the relaxation of spin density or other spin-related variables, we leave the research work to future.
  
\section*{Acknowledgments}
J.H. is grateful to Jiaxing Zhao and Ziyue Wang for reading the script and helpful advice.  This work was supported by the NSFC Grant No.11890710, No.11890712 and No.12035006.
\clearpage
\begin{appendix}
	
\section{Thermodynamic  Integral} \label{int}
In this section, we calculate thermodynamic  integrals used in main text. The first one is
\begin{align}
\label{I1}
I^{\alpha_1\cdots\alpha_n}&\equiv 2\int \frac{\rm dP}{(2\pi)^3 }\,p^{\alpha_1}p^{\alpha_2}\cdots p^{\alpha_n}e^{\xi-\beta \cdot p}\nn\\
&=I_{n0}u^{\alpha_1}\cdots u^{\alpha_n}+I_{n1}(\Delta^{\alpha_1\alpha_2}u^{\alpha_3\cdots\alpha_n}+\text{permutations})+\cdots,
\end{align}
where in the second equality we have employed the analysis of Lorentz covariance. By  projecting  $u^\alpha$ and $\Delta^{\alpha\beta}$ onto Eq.(\ref{I1}), the scalar coefficients are defined as
\begin{eqnarray}
\label{Inq}
I_{nq} &\equiv& \frac{2}{(2q+1)!!} \int \frac{\rm dP}{(2\pi)^3}\,(u\cdot p)^{n-2q} (\Delta_{\alpha\beta} p^{\alpha} p^{\beta})^q e^{\xi-\beta \cdot p},
\end{eqnarray}
with $K_n(z)$ representing the modified Bessel functions of the second kind 
\begin{eqnarray}
K_n(z) &\equiv& \int_0^{\infty} \mathrm{d}x\, \cosh(nx)\, e^{- z \cosh x}.
\end{eqnarray}
Especially, we have $I_{10}=n$, $I_{20}=e$ , $I_{31}=-h$ and 
\begin{align}
I_{30}(z)=\frac{T^5z^5}{32\pi^2}\big(K_5(z)+K_3(z)-2K_1(z)\big),
\end{align}
with $z\equiv\frac{m}{T}$.

Additionally,  the following similar formulas are also of great use,
\begin{align}
L^{\alpha_1\cdots\alpha_n}&\equiv 2\int \frac{\rm dP}{(2\pi)^3 (p^0+m)}\,p^{\alpha_1}p^{\alpha_2}\cdots p^{\alpha_n}e^{\xi-\beta \cdot p}\nn\\
&=L_{n0}u^{\alpha_1}\cdots u^{\alpha_n}+L_{n1}(\Delta^{\alpha_1\alpha_2}u^{\alpha_3\cdots\alpha_n}+\text{permutations})+\cdots,\nn\\
N^{\alpha_1\cdots\alpha_n}&\equiv 2\int \frac{\rm dP}{(2\pi)^3 (p^0+m)^2}\,p^{\alpha_1}p^{\alpha_2}\cdots p^{\alpha_n}e^{\xi-\beta \cdot p}\nn\\
&=N_{n0}u^{\alpha_1}\cdots u^{\alpha_n}+N_{n1}(\Delta^{\alpha_1\alpha_2}u^{\alpha_3\cdots\alpha_n}+\text{permutations})+\cdots .
\end{align}
Similarly,  $L_{nq}$ and $N_{nq}$ are  also obtained like Eq.(\ref{Inq}). Generally speaking, these scalar integrals may not be analytically worked by  expressing them with the modified Bessel functions of the second kind $K_n(z)$ due to extra factor appearing in the integrations and we need to turn to numerical integration.

\section{Normalized factors} \label{int1}
In the main text, we utilize two auxiliary unit vectors $j^\mu$ and $v^\mu$, which form a triad with the vectors $u$ and $l$,
\begin{align}
&u\cdot l=u\cdot j=u\cdot v=l\cdot j=l\cdot v=j\cdot v=0,\nn\\
&l^2=j^2=v^2=-1.
\end{align}
Therefore  $p^\mu$ and $J^{\mu\nu}$ can be readily expanded as
\begin{align}
&p^\mu=u\cdot p \,u^\mu+l\cdot p \,l^\mu+j\cdot p \,j^\mu+v\cdot p\, v^\mu,\nn\\
&J^{\mu\nu}=u_\mu J^{\mu\nu}l_\nu-l_\mu J^{\mu\nu}u_\nu+u_\mu J^{\mu\nu}j_\nu-j_\mu J^{\mu\nu}u_\nu+u_\mu J^{\mu\nu}v_\nu-v_\mu J^{\mu\nu}u_\nu\nn\\
&\quad\,\,\,\,+l_\mu J^{\mu\nu}j_\nu-j_\mu J^{\mu\nu}l_\nu+l_\mu J^{\mu\nu}v_\nu-v_\mu J^{\mu\nu}l_\nu+j_\mu J^{\mu\nu}v_\nu-v_\mu J^{\mu\nu}j_\nu.
\end{align}
As  the total angular momentum $J^{\mu\nu}$ is antisymmetric, the effective basis set are $(1,u\cdot p,l\cdot p,j\cdot p,v\cdot p, u_\mu J^{\mu\nu}l_\nu,u_\mu J^{\mu\nu}j_\nu,\nn\\u_\mu J^{\mu\nu}v_\nu,l_\mu J^{\mu\nu}j_\nu,l_\mu J^{\mu\nu}v_\nu,j_\mu J^{\mu\nu}v_\nu)$. Throughout the paper, they are ordered like this way and are labeled by the $i$-th basis respectively ($i=1,2,\cdots,11$).

Then the normalized factors appearing in Eq.(\ref{sets})  are 
\begin{eqnarray}
\label{Inq3}
V_{1,1}&=&\exp(\xi)\int \frac{\rm d\Gamma}{(2\pi)^3}\frac{u\cdot p}{T}\exp(-\beta \cdot p)=\frac{n}{T},\nn\\
\quad V_{2,2}&=&\exp(\xi)\int \frac{\rm d\Gamma}{(2\pi)^3}\frac{(u\cdot p-\frac{e}{n})^2(u\cdot p)}{T^3}\exp(-\beta \cdot p)=\frac{I_{30}-\frac{e^2}{n}}{T^3},\nn\\
V_{3,3}&=&V_{4,4}=V_{5,5}=\exp(\xi)\int \frac{\rm d\Gamma}{(2\pi)^3}\frac{(u\cdot p)(l\cdot p)^2}{T^3}\exp(-\beta \cdot p)=\frac{h}{T^2}.
\end{eqnarray}

When calculating  the remaining normalized factors, recall the calibration settings mentioned after Eq.(\ref{colleqq1}) $\hat{t}=u=(1,0,0,0)$, and we obtain
\begin{align}
V_{6,6}=V_{7,7}=V_{8,8}&=\exp(\xi)\int \frac{\rm d\Gamma}{(2\pi)^3}\frac{u\cdot p}{T}u_\mu J^{\mu\nu}l_\nu u_\rho J^{\rho\sigma}l_\sigma\exp(-\beta \cdot p)\nn\\
&=\frac{1}{2m^2T}(-I_{31}+2L_{41}-N_{51}),\nn\\
V_{9,9}=V_{10,10}=V_{11,11}&=\exp(\xi)\int \frac{\rm d\Gamma}{(2\pi)^3}l_\mu J^{\mu\nu}j_\nu(p\cdot u) l_\rho J^{\rho\sigma}j_\sigma\exp(-\beta \cdot p)\nn\\
&=\frac{I_{30}+I_{31}+4L_{41}+10N_{52}}{4m^2T},
\end{align}
where $L_{nq}$ and $N_{nq}$ are defined in Appendix.\ref{int} and  the factors $V_{2,2},V_{6,6}$ and $V_{9,9}$ are  positive.

\section{Perturbation Matrix Elements } \label{pertu}
We compute the perturbation matrix elements used in Eq.(\ref{secu}). The matrix elements sandwiching the $p\cdot\kappa$ term are written as
\begin{eqnarray}
\label{pk}
H_{1,3}&=&\frac{\exp(\xi)}{\sqrt{V_{1,1}V_{3,3}}T^2}\int \frac{\rm d\Gamma}{(2\pi)^3}\kappa\cdot p\,l\cdot p\exp(-\beta \cdot p)=\frac{P\kappa}{\sqrt{V_{1,1}V_{3,3}}T^2},\nn\\
H_{2,3}&=&\frac{\exp(\xi)}{\sqrt{V_{2,2}V_{3,3}}T^3}\int \frac{\rm d\Gamma}{(2\pi)^3}(u\cdot p-\frac{e}{n})p\cdot\kappa(l\cdot p)\exp(-\beta \cdot p)=\frac{P\kappa}{\sqrt{V_{2,2}V_{3,3}}T^2},\nn\\
H_{7,9}=H_{8,10}&=&\frac{\exp(\xi)}{\sqrt{V_{6,6}V_{9,9}}}\int \frac{\rm d\Gamma}{(2\pi)^3}\frac{p\cdot\kappa}{T}u_\mu J^{\mu\nu}j_\nu l_\rho J^{\rho\sigma}j_\sigma\exp(-\beta \cdot p)=\frac{(-I_{31}+L_{41}-5L_{42}+5N_{52})\kappa}{4m^2T\sqrt{V_{6,6}V_{9,9}}}.
\end{eqnarray}
Noticing that the perturbation matrix owns a symmetry of transposition $H_{i,j}=H_{j,i}$. Except the above matrix elements and their transpositions, the others are all zero. 
\clearpage
\section{Proof  of hermiticity of linearized collision operator} \label{prove}
First, we note the linearized collision operator to be proved is 
\begin{eqnarray}
\mathcal{L}_{1}[\phi]
&\equiv &\frac{1}{(2\pi)^3} \int d\Gamma^\prime d\Gamma_1 d\Gamma_2  \,
\bar{\mathcal{W}}\,F[p,p^\prime,p_1,p_2;\bm{s},\bm{s}^\prime,\bm{s}_1,\bm{s}_2] \,  \exp(\xi-u\cdot p^\prime\,)\nn\\
&\times& \Big[\phi(x,p,\bm{s})+\phi(x,p^\prime,\bm{s}^\prime)-\phi(x,p_1,\bm{s}_1)-\phi(x,p_2,\bm{s}_2)\,\Big].
\label{colleqqcl2}
\end{eqnarray}

In the sense of the definition of inner product Eq.(\ref{inner}), we obtain,
\begin{eqnarray}\label{prove1}
(\mathcal{L}_{1}[\phi],\psi)
&= &\frac{1}{(2\pi)^6} \int d\Gamma d\Gamma^\prime d\Gamma_1 d\Gamma_2  \,
\bar{\mathcal{W}}\, F[p,p^\prime,p_1,p_2;\bm{s},\bm{s}^\prime,\bm{s}_1,\bm{s}_2]\,  \exp(2\xi-u\cdot (p+p^\prime)\,)\nn\\
&\times& \Big[\phi(x,p,\bm{s})+\phi(x,p^\prime,\bm{s}^\prime)-\phi(x,p_1,\bm{s}_1)-\phi(x,p_2,\bm{s}_2)\,\Big]\psi(x,p,\bm{s})\nn\\
&= &\frac{1}{2(2\pi)^6} \int d\Gamma d\Gamma^\prime d\Gamma_1 d\Gamma_2  \,
\bar{\mathcal{W}} \,F[p,p^\prime,p_1,p_2;\bm{s},\bm{s}^\prime,\bm{s}_1,\bm{s}_2]\,  \exp(2\xi-u\cdot (p+p^\prime)\,)\nn\\
&\times& \Big[\phi(x,p,\bm{s})+\phi(x,p^\prime,\bm{s}^\prime)-\phi(x,p_1,\bm{s}_1)-\phi(x,p_2,\bm{s}_2)\,\Big](\psi(x,p,\bm{s})+\psi(x,p^\prime,\bm{s}^\prime))\nn\\
&= &\frac{1}{4(2\pi)^6} \int d\Gamma d\Gamma^\prime d\Gamma_1 d\Gamma_2  \,
\bar{\mathcal{W}} \,F[p,p^\prime,p_1,p_2;\bm{s},\bm{s}^\prime,\bm{s}_1,\bm{s}_2]\,  \exp(2\xi-u\cdot (p+p^\prime)\,)\nn\\
&\times& \Big[\big(\phi(x,p,\bm{s})+\phi(x,p^\prime,\bm{s}^\prime)\,\big)\big(\psi(x,p,\bm{s})+\psi(x,p^\prime,\bm{s}^\prime)\,\big)-2\big(\phi(x,p_1,\bm{s}_1)+\phi(x,p_2,\bm{s}_2)\,\big)\big(\psi(x,p,\bm{s})+\psi(x,p^\prime,\bm{s}^\prime)\,\big)\,\nn\\
&+&\big(\phi(x,p_1,\bm{s}_1)+\phi(x,p_2,\bm{s}_2)\,\big)\big(\psi(x,p_1,\bm{s}_1)+\psi(x,p_2,\bm{s}_2)\,\big)\Big].
\end{eqnarray}
Only by neglecting the polarized part in collision rate can the third step be implemented.  To see this, when interchanging $(p,\bm{s})$ with $(p_1,\bm{s}_1)$ (for simplicity we omit $(p^\prime,\bm{s}^\prime)$ and $(p_2,\bm{s}_2)$ because of duplication and other trivial factors)  and taking  the full $h_{sr}$ into account,
\begin{align}
\label{lhs}
\mathcal{W}&=\sum_{s,r,s_1,r_1} h_{s r} (p,\bm{s})  
h_{s_1 r_1}(p_1,\bm{s}_1)  \langle{p,r|t|p_1,s_1}\rangle
\langle{p_1,r_1|t^\dagger|p,s}\rangle\\
\label{rhs}
\mathcal{W}\Big[(p,\bm{s}) \leftrightarrow (p_1,\bm{s}_1)\Big]&=\sum_{s,r,s_1,r_1} h_{s r} (p_1,\bm{s}_1)  
h_{s_1 r_1}(p,\bm{s})  \langle{p_1,r|t|p,s_1}\rangle
\langle{p,r_1|t^\dagger|p_1,s}\rangle\nn\\
&=\sum_{s,r,s_1,r_1} h_{s r} (p,\bm{s})  
h_{s_1 r_1}(p_1,\bm{s}_1)  \langle{p_1,r_1|t|p,s}\rangle
\langle{p,r|t^\dagger|p_1,s_1}\rangle.	
\end{align}
In general, Eq.(\ref{lhs}) is not equivalent to Eq.(\ref{rhs}) unless $t$ is Hermitian. From Eq.(\ref{prove1}), we conclude that $(\mathcal{L}_{1}[\phi],\phi)\geq 0$ if the transition rate respects this exchanging symmetry.
From the above argument, we find the non-negative property of linearized collision operator lies in the negligence of the polarized part of collision rate $\mathcal{W}$. When neglecting the polarized vector $n_{sr}$, the collision rate can be cast into the traditional form of  cross section and thus symmetric for exchanging the momentums of initial state and final state, which is called the principle of detailed balance.

Continuing the discussion following Eq.(\ref{prove1}) with the collision rate replaced by the reduced (unpolarized) one, we find with assistance of the detailed balance,
\begin{align}
\label{ls}
(\mathcal{L}_{1}[\phi],\psi)=(\mathcal{L}_{1}[\psi],\phi)=(\phi,\mathcal{L}_{1}[\psi]).
\end{align}	
To prove this,  interchanging $(p,\bm{s}), (p^\prime,\bm{s}^\prime)$ with $(p_1,\bm{s}_1), (p_2,\bm{s}_2)$
\begin{align}
\big(\phi(x,p_1,\bm{s}_1)+\phi(x,p_2,\bm{s}_2)\,\big)\big(\psi(x,p,\bm{s})+\psi(x,p^\prime,\bm{s}^\prime)\,\big)\rightarrow\big(\psi(x,p_1,\bm{s}_1)+\psi(x,p_2,\bm{s}_2)\,\big)\big(\phi(x,p,\bm{s})+\phi(x,p^\prime,\bm{s}^\prime)\,\big),
\end{align}
the integral in Eq.(\ref{prove1}) over this term should be invariant. Such property
is employed  for the second term within the bracket  of Eq.(\ref{prove1}) invoking the symmetry property of the transition rate (the other terms have this exchange symmetry of $\phi$ and $\psi$ already). Without detailed balance,  Eq.(\ref{ls}) can not hold, thus the polarized part of collision rate is also responsible for non-hermiticity of linearized collision term. To conclude, we derive the non-negative self-adjoint property for the unpolarized linearized collision operator.
\clearpage

\end{appendix}
\bibliographystyle{apsrev}
\bibliography{spinmode}{}

\begin{thebibliography}{44}
\expandafter\ifx\csname natexlab\endcsname\relax\def\natexlab#1{#1}\fi
\expandafter\ifx\csname bibnamefont\endcsname\relax
  \def\bibnamefont#1{#1}\fi
\expandafter\ifx\csname bibfnamefont\endcsname\relax
  \def\bibfnamefont#1{#1}\fi
\expandafter\ifx\csname citenamefont\endcsname\relax
  \def\citenamefont#1{#1}\fi
\expandafter\ifx\csname url\endcsname\relax
  \def\url#1{\texttt{#1}}\fi
\expandafter\ifx\csname urlprefix\endcsname\relax\def\urlprefix{URL }\fi
\providecommand{\bibinfo}[2]{#2}
\providecommand{\eprint}[2][]{\url{#2}}

\bibitem[{\citenamefont{Weickgenannt
  et~al.}(2021{\natexlab{a}})\citenamefont{Weickgenannt, Speranza, Sheng, Wang,
  and Rischke}}]{Weickgenannt:2021cuo}
\bibinfo{author}{\bibfnamefont{N.}~\bibnamefont{Weickgenannt}},
  \bibinfo{author}{\bibfnamefont{E.}~\bibnamefont{Speranza}},
  \bibinfo{author}{\bibfnamefont{X.-l.} \bibnamefont{Sheng}},
  \bibinfo{author}{\bibfnamefont{Q.}~\bibnamefont{Wang}}, \bibnamefont{and}
  \bibinfo{author}{\bibfnamefont{D.~H.} \bibnamefont{Rischke}},
  \bibinfo{journal}{Phys. Rev. D} \textbf{\bibinfo{volume}{104}},
  \bibinfo{pages}{016022} (\bibinfo{year}{2021}{\natexlab{a}}),
  \eprint{2103.04896}.

\bibitem[{\citenamefont{Adamczyk et~al.}(2017)}]{STAR:2017ckg}
\bibinfo{author}{\bibfnamefont{L.}~\bibnamefont{Adamczyk}} \bibnamefont{et~al.}
  (\bibinfo{collaboration}{STAR}), \bibinfo{journal}{Nature}
  \textbf{\bibinfo{volume}{548}}, \bibinfo{pages}{62} (\bibinfo{year}{2017}),
  \eprint{1701.06657}.

\bibitem[{\citenamefont{Alpatov}(2020)}]{Alpatov:2020iev}
\bibinfo{author}{\bibfnamefont{E.}~\bibnamefont{Alpatov}}
  (\bibinfo{collaboration}{), STAR (for the}), \bibinfo{journal}{J. Phys. Conf.
  Ser.} \textbf{\bibinfo{volume}{1690}}, \bibinfo{pages}{012120}
  (\bibinfo{year}{2020}).

\bibitem[{\citenamefont{Wei et~al.}(2019)\citenamefont{Wei, Deng, and
  Huang}}]{Wei:2018zfb}
\bibinfo{author}{\bibfnamefont{D.-X.} \bibnamefont{Wei}},
  \bibinfo{author}{\bibfnamefont{W.-T.} \bibnamefont{Deng}}, \bibnamefont{and}
  \bibinfo{author}{\bibfnamefont{X.-G.} \bibnamefont{Huang}},
  \bibinfo{journal}{Phys. Rev. C} \textbf{\bibinfo{volume}{99}},
  \bibinfo{pages}{014905} (\bibinfo{year}{2019}), \eprint{1810.00151}.

\bibitem[{\citenamefont{Karpenko and Becattini}(2017)}]{Karpenko:2016jyx}
\bibinfo{author}{\bibfnamefont{I.}~\bibnamefont{Karpenko}} \bibnamefont{and}
  \bibinfo{author}{\bibfnamefont{F.}~\bibnamefont{Becattini}},
  \bibinfo{journal}{Eur. Phys. J. C} \textbf{\bibinfo{volume}{77}},
  \bibinfo{pages}{213} (\bibinfo{year}{2017}), \eprint{1610.04717}.

\bibitem[{\citenamefont{Csernai et~al.}(2019)\citenamefont{Csernai, Kapusta,
  and Welle}}]{Csernai:2018yok}
\bibinfo{author}{\bibfnamefont{L.}~\bibnamefont{Csernai}},
  \bibinfo{author}{\bibfnamefont{J.}~\bibnamefont{Kapusta}}, \bibnamefont{and}
  \bibinfo{author}{\bibfnamefont{T.}~\bibnamefont{Welle}},
  \bibinfo{journal}{Phys. Rev. C} \textbf{\bibinfo{volume}{99}},
  \bibinfo{pages}{021901} (\bibinfo{year}{2019}), \eprint{1807.11521}.

\bibitem[{\citenamefont{Li et~al.}(2017)\citenamefont{Li, Pang, Wang, and
  Xia}}]{Li:2017slc}
\bibinfo{author}{\bibfnamefont{H.}~\bibnamefont{Li}},
  \bibinfo{author}{\bibfnamefont{L.-G.} \bibnamefont{Pang}},
  \bibinfo{author}{\bibfnamefont{Q.}~\bibnamefont{Wang}}, \bibnamefont{and}
  \bibinfo{author}{\bibfnamefont{X.-L.} \bibnamefont{Xia}},
  \bibinfo{journal}{Phys. Rev. C} \textbf{\bibinfo{volume}{96}},
  \bibinfo{pages}{054908} (\bibinfo{year}{2017}), \eprint{1704.01507}.

\bibitem[{\citenamefont{Bzdak}(2017)}]{Bzdak:2017shg}
\bibinfo{author}{\bibfnamefont{A.}~\bibnamefont{Bzdak}},
  \bibinfo{journal}{Phys. Rev. D} \textbf{\bibinfo{volume}{96}},
  \bibinfo{pages}{056011} (\bibinfo{year}{2017}), \eprint{1703.03003}.

\bibitem[{\citenamefont{Shi et~al.}(2019)\citenamefont{Shi, Li, and
  Liao}}]{Shi:2017wpk}
\bibinfo{author}{\bibfnamefont{S.}~\bibnamefont{Shi}},
  \bibinfo{author}{\bibfnamefont{K.}~\bibnamefont{Li}}, \bibnamefont{and}
  \bibinfo{author}{\bibfnamefont{J.}~\bibnamefont{Liao}},
  \bibinfo{journal}{Phys. Lett. B} \textbf{\bibinfo{volume}{788}},
  \bibinfo{pages}{409} (\bibinfo{year}{2019}), \eprint{1712.00878}.

\bibitem[{\citenamefont{Sun and Ko}(2017)}]{Sun:2017xhx}
\bibinfo{author}{\bibfnamefont{Y.}~\bibnamefont{Sun}} \bibnamefont{and}
  \bibinfo{author}{\bibfnamefont{C.~M.} \bibnamefont{Ko}},
  \bibinfo{journal}{Phys. Rev. C} \textbf{\bibinfo{volume}{96}},
  \bibinfo{pages}{024906} (\bibinfo{year}{2017}), \eprint{1706.09467}.

\bibitem[{\citenamefont{Ivanov et~al.}(2020)\citenamefont{Ivanov, Toneev, and
  Soldatov}}]{Ivanov:2019wzg}
\bibinfo{author}{\bibfnamefont{Y.~B.} \bibnamefont{Ivanov}},
  \bibinfo{author}{\bibfnamefont{V.~D.} \bibnamefont{Toneev}},
  \bibnamefont{and} \bibinfo{author}{\bibfnamefont{A.~A.}
  \bibnamefont{Soldatov}}, \bibinfo{journal}{Phys. Atom. Nucl.}
  \textbf{\bibinfo{volume}{83}}, \bibinfo{pages}{179} (\bibinfo{year}{2020}),
  \eprint{1910.01332}.

\bibitem[{\citenamefont{Xie et~al.}(2017)\citenamefont{Xie, Wang, and
  Csernai}}]{Xie:2017upb}
\bibinfo{author}{\bibfnamefont{Y.}~\bibnamefont{Xie}},
  \bibinfo{author}{\bibfnamefont{D.}~\bibnamefont{Wang}}, \bibnamefont{and}
  \bibinfo{author}{\bibfnamefont{L.~P.} \bibnamefont{Csernai}},
  \bibinfo{journal}{Phys. Rev. C} \textbf{\bibinfo{volume}{95}},
  \bibinfo{pages}{031901} (\bibinfo{year}{2017}), \eprint{1703.03770}.

\bibitem[{\citenamefont{Adam et~al.}(2019)}]{Adam:2019srw}
\bibinfo{author}{\bibfnamefont{J.}~\bibnamefont{Adam}} \bibnamefont{et~al.}
  (\bibinfo{collaboration}{STAR}), \bibinfo{journal}{Phys. Rev. Lett.}
  \textbf{\bibinfo{volume}{123}}, \bibinfo{pages}{132301}
  (\bibinfo{year}{2019}), \eprint{1905.11917}.

\bibitem[{\citenamefont{Adam et~al.}(2018)}]{Adam:2018ivw}
\bibinfo{author}{\bibfnamefont{J.}~\bibnamefont{Adam}} \bibnamefont{et~al.}
  (\bibinfo{collaboration}{STAR}), \bibinfo{journal}{Phys. Rev. C}
  \textbf{\bibinfo{volume}{98}}, \bibinfo{pages}{014910}
  (\bibinfo{year}{2018}), \eprint{1805.04400}.

\bibitem[{\citenamefont{Becattini and Karpenko}(2018)}]{Becattini:2017gcx}
\bibinfo{author}{\bibfnamefont{F.}~\bibnamefont{Becattini}} \bibnamefont{and}
  \bibinfo{author}{\bibfnamefont{I.}~\bibnamefont{Karpenko}},
  \bibinfo{journal}{Phys. Rev. Lett.} \textbf{\bibinfo{volume}{120}},
  \bibinfo{pages}{012302} (\bibinfo{year}{2018}), \eprint{1707.07984}.

\bibitem[{\citenamefont{Xia et~al.}(2018)\citenamefont{Xia, Li, Tang, and
  Wang}}]{Xia:2018tes}
\bibinfo{author}{\bibfnamefont{X.-L.} \bibnamefont{Xia}},
  \bibinfo{author}{\bibfnamefont{H.}~\bibnamefont{Li}},
  \bibinfo{author}{\bibfnamefont{Z.-B.} \bibnamefont{Tang}}, \bibnamefont{and}
  \bibinfo{author}{\bibfnamefont{Q.}~\bibnamefont{Wang}},
  \bibinfo{journal}{Phys. Rev. C} \textbf{\bibinfo{volume}{98}},
  \bibinfo{pages}{024905} (\bibinfo{year}{2018}), \eprint{1803.00867}.

\bibitem[{\citenamefont{Florkowski et~al.}(2018)\citenamefont{Florkowski,
  Friman, Jaiswal, and Speranza}}]{Florkowski:2017ruc}
\bibinfo{author}{\bibfnamefont{W.}~\bibnamefont{Florkowski}},
  \bibinfo{author}{\bibfnamefont{B.}~\bibnamefont{Friman}},
  \bibinfo{author}{\bibfnamefont{A.}~\bibnamefont{Jaiswal}}, \bibnamefont{and}
  \bibinfo{author}{\bibfnamefont{E.}~\bibnamefont{Speranza}},
  \bibinfo{journal}{Phys. Rev. C} \textbf{\bibinfo{volume}{97}},
  \bibinfo{pages}{041901} (\bibinfo{year}{2018}), \eprint{1705.00587}.

\bibitem[{\citenamefont{Peng et~al.}(2021)\citenamefont{Peng, Zhang, Sheng, and
  Wang}}]{Peng:2021ago}
\bibinfo{author}{\bibfnamefont{H.-H.} \bibnamefont{Peng}},
  \bibinfo{author}{\bibfnamefont{J.-J.} \bibnamefont{Zhang}},
  \bibinfo{author}{\bibfnamefont{X.-L.} \bibnamefont{Sheng}}, \bibnamefont{and}
  \bibinfo{author}{\bibfnamefont{Q.}~\bibnamefont{Wang}},
  \bibinfo{journal}{Chin. Phys. Lett.} \textbf{\bibinfo{volume}{38}},
  \bibinfo{pages}{116701} (\bibinfo{year}{2021}), \eprint{2107.00448}.

\bibitem[{\citenamefont{Becattini and Tinti}(2010)}]{Becattini:2009wh}
\bibinfo{author}{\bibfnamefont{F.}~\bibnamefont{Becattini}} \bibnamefont{and}
  \bibinfo{author}{\bibfnamefont{L.}~\bibnamefont{Tinti}},
  \bibinfo{journal}{Annals Phys.} \textbf{\bibinfo{volume}{325}},
  \bibinfo{pages}{1566} (\bibinfo{year}{2010}), \eprint{0911.0864}.

\bibitem[{\citenamefont{Florkowski et~al.}(2019)\citenamefont{Florkowski,
  Kumar, and Ryblewski}}]{Florkowski:2018fap}
\bibinfo{author}{\bibfnamefont{W.}~\bibnamefont{Florkowski}},
  \bibinfo{author}{\bibfnamefont{A.}~\bibnamefont{Kumar}}, \bibnamefont{and}
  \bibinfo{author}{\bibfnamefont{R.}~\bibnamefont{Ryblewski}},
  \bibinfo{journal}{Prog. Part. Nucl. Phys.} \textbf{\bibinfo{volume}{108}},
  \bibinfo{pages}{103709} (\bibinfo{year}{2019}), \eprint{1811.04409}.

\bibitem[{\citenamefont{Hattori et~al.}(2019)\citenamefont{Hattori, Hongo,
  Huang, Matsuo, and Taya}}]{Hattori:2019lfp}
\bibinfo{author}{\bibfnamefont{K.}~\bibnamefont{Hattori}},
  \bibinfo{author}{\bibfnamefont{M.}~\bibnamefont{Hongo}},
  \bibinfo{author}{\bibfnamefont{X.-G.} \bibnamefont{Huang}},
  \bibinfo{author}{\bibfnamefont{M.}~\bibnamefont{Matsuo}}, \bibnamefont{and}
  \bibinfo{author}{\bibfnamefont{H.}~\bibnamefont{Taya}},
  \bibinfo{journal}{Phys. Lett. B} \textbf{\bibinfo{volume}{795}},
  \bibinfo{pages}{100} (\bibinfo{year}{2019}), \eprint{1901.06615}.

\bibitem[{\citenamefont{Bhadury
  et~al.}(2021{\natexlab{a}})\citenamefont{Bhadury, Florkowski, Jaiswal, Kumar,
  and Ryblewski}}]{Bhadury:2020cop}
\bibinfo{author}{\bibfnamefont{S.}~\bibnamefont{Bhadury}},
  \bibinfo{author}{\bibfnamefont{W.}~\bibnamefont{Florkowski}},
  \bibinfo{author}{\bibfnamefont{A.}~\bibnamefont{Jaiswal}},
  \bibinfo{author}{\bibfnamefont{A.}~\bibnamefont{Kumar}}, \bibnamefont{and}
  \bibinfo{author}{\bibfnamefont{R.}~\bibnamefont{Ryblewski}},
  \bibinfo{journal}{Phys. Rev. D} \textbf{\bibinfo{volume}{103}},
  \bibinfo{pages}{014030} (\bibinfo{year}{2021}{\natexlab{a}}),
  \eprint{2008.10976}.

\bibitem[{\citenamefont{Fukushima and Pu}(2021)}]{Fukushima:2020ucl}
\bibinfo{author}{\bibfnamefont{K.}~\bibnamefont{Fukushima}} \bibnamefont{and}
  \bibinfo{author}{\bibfnamefont{S.}~\bibnamefont{Pu}}, \bibinfo{journal}{Phys.
  Lett. B} \textbf{\bibinfo{volume}{817}}, \bibinfo{pages}{136346}
  (\bibinfo{year}{2021}), \eprint{2010.01608}.

\bibitem[{\citenamefont{Hu}(2021{\natexlab{a}})}]{Hu:2021lnx}
\bibinfo{author}{\bibfnamefont{J.}~\bibnamefont{Hu}}, \bibinfo{journal}{Phys.
  Rev. D} \textbf{\bibinfo{volume}{103}}, \bibinfo{pages}{116015}
  (\bibinfo{year}{2021}{\natexlab{a}}), \eprint{2101.08440}.

\bibitem[{\citenamefont{Hu}(2022{\natexlab{a}})}]{Hu:2021pwh}
\bibinfo{author}{\bibfnamefont{J.}~\bibnamefont{Hu}}, \bibinfo{journal}{Phys.
  Rev. D} \textbf{\bibinfo{volume}{105}}, \bibinfo{pages}{076009}
  (\bibinfo{year}{2022}{\natexlab{a}}), \eprint{2111.03571}.

\bibitem[{\citenamefont{Hu}(2022{\natexlab{b}})}]{Hu:2022xjn}
\bibinfo{author}{\bibfnamefont{J.}~\bibnamefont{Hu}}, \bibinfo{journal}{Phys.
  Rev. D} \textbf{\bibinfo{volume}{105}}, \bibinfo{pages}{096021}
  (\bibinfo{year}{2022}{\natexlab{b}}), \eprint{2204.12946}.

\bibitem[{\citenamefont{Hu and Xu}(2022)}]{Hu:2022mvl}
\bibinfo{author}{\bibfnamefont{J.}~\bibnamefont{Hu}} \bibnamefont{and}
  \bibinfo{author}{\bibfnamefont{Z.}~\bibnamefont{Xu}} (\bibinfo{year}{2022}),
  \eprint{2205.15755}.

\bibitem[{\citenamefont{Weickgenannt
  et~al.}(2021{\natexlab{b}})\citenamefont{Weickgenannt, Speranza, Sheng, Wang,
  and Rischke}}]{Weickgenannt:2020aaf}
\bibinfo{author}{\bibfnamefont{N.}~\bibnamefont{Weickgenannt}},
  \bibinfo{author}{\bibfnamefont{E.}~\bibnamefont{Speranza}},
  \bibinfo{author}{\bibfnamefont{X.-l.} \bibnamefont{Sheng}},
  \bibinfo{author}{\bibfnamefont{Q.}~\bibnamefont{Wang}}, \bibnamefont{and}
  \bibinfo{author}{\bibfnamefont{D.~H.} \bibnamefont{Rischke}},
  \bibinfo{journal}{Phys. Rev. Lett.} \textbf{\bibinfo{volume}{127}},
  \bibinfo{pages}{052301} (\bibinfo{year}{2021}{\natexlab{b}}),
  \eprint{2005.01506}.

\bibitem[{\citenamefont{Yang et~al.}(2020)\citenamefont{Yang, Hattori, and
  Hidaka}}]{Yang:2020hri}
\bibinfo{author}{\bibfnamefont{D.-L.} \bibnamefont{Yang}},
  \bibinfo{author}{\bibfnamefont{K.}~\bibnamefont{Hattori}}, \bibnamefont{and}
  \bibinfo{author}{\bibfnamefont{Y.}~\bibnamefont{Hidaka}},
  \bibinfo{journal}{JHEP} \textbf{\bibinfo{volume}{20}}, \bibinfo{pages}{070}
  (\bibinfo{year}{2020}), \eprint{2002.02612}.

\bibitem[{\citenamefont{Wang et~al.}(2021)\citenamefont{Wang, Guo, and
  Zhuang}}]{Wang:2020pej}
\bibinfo{author}{\bibfnamefont{Z.}~\bibnamefont{Wang}},
  \bibinfo{author}{\bibfnamefont{X.}~\bibnamefont{Guo}}, \bibnamefont{and}
  \bibinfo{author}{\bibfnamefont{P.}~\bibnamefont{Zhuang}},
  \bibinfo{journal}{Eur. Phys. J. C} \textbf{\bibinfo{volume}{81}},
  \bibinfo{pages}{799} (\bibinfo{year}{2021}), \eprint{2009.10930}.

\bibitem[{\citenamefont{Sheng et~al.}(2021)\citenamefont{Sheng, Weickgenannt,
  Speranza, Rischke, and Wang}}]{Sheng:2021kfc}
\bibinfo{author}{\bibfnamefont{X.-L.} \bibnamefont{Sheng}},
  \bibinfo{author}{\bibfnamefont{N.}~\bibnamefont{Weickgenannt}},
  \bibinfo{author}{\bibfnamefont{E.}~\bibnamefont{Speranza}},
  \bibinfo{author}{\bibfnamefont{D.~H.} \bibnamefont{Rischke}},
  \bibnamefont{and} \bibinfo{author}{\bibfnamefont{Q.}~\bibnamefont{Wang}},
  \bibinfo{journal}{Phys. Rev. D} \textbf{\bibinfo{volume}{104}},
  \bibinfo{pages}{016029} (\bibinfo{year}{2021}), \eprint{2103.10636}.

\bibitem[{\citenamefont{Weickgenannt
  et~al.}(2022{\natexlab{a}})\citenamefont{Weickgenannt, Wagner, Speranza, and
  Rischke}}]{Weickgenannt:2022zxs}
\bibinfo{author}{\bibfnamefont{N.}~\bibnamefont{Weickgenannt}},
  \bibinfo{author}{\bibfnamefont{D.}~\bibnamefont{Wagner}},
  \bibinfo{author}{\bibfnamefont{E.}~\bibnamefont{Speranza}}, \bibnamefont{and}
  \bibinfo{author}{\bibfnamefont{D.}~\bibnamefont{Rischke}}
  (\bibinfo{year}{2022}{\natexlab{a}}), \eprint{2203.04766}.

\bibitem[{\citenamefont{Balescu}(1975)}]{Balescu}
\bibinfo{author}{\bibfnamefont{R.}~\bibnamefont{Balescu}},
  \emph{\bibinfo{title}{Equilibrium and Nonequilibrium Statistical Mechanics}}
  (\bibinfo{publisher}{A Wiley-Interscience}, \bibinfo{year}{1975}).

\bibitem[{\citenamefont{R\'esibois and Leener}(1977)}]{Resi}
\bibinfo{author}{\bibfnamefont{P.}~\bibnamefont{R\'esibois}} \bibnamefont{and}
  \bibinfo{author}{\bibfnamefont{M.~D.} \bibnamefont{Leener}},
  \emph{\bibinfo{title}{Classical Kinetic Theory of Fluids}}
  (\bibinfo{publisher}{A Wiley-Interscience}, \bibinfo{year}{1977}).

\bibitem[{\citenamefont{Zamanian et~al.}(2010)\citenamefont{Zamanian, Marklund,
  and Brodin}}]{Zamanian:2010zz}
\bibinfo{author}{\bibfnamefont{J.}~\bibnamefont{Zamanian}},
  \bibinfo{author}{\bibfnamefont{M.}~\bibnamefont{Marklund}}, \bibnamefont{and}
  \bibinfo{author}{\bibfnamefont{G.}~\bibnamefont{Brodin}},
  \bibinfo{journal}{New J. Phys.} \textbf{\bibinfo{volume}{12}},
  \bibinfo{pages}{043019} (\bibinfo{year}{2010}).

\bibitem[{\citenamefont{Ekman et~al.}(2017)\citenamefont{Ekman, Asenjo, and
  Zamanian}}]{Ekman:2017kxi}
\bibinfo{author}{\bibfnamefont{R.}~\bibnamefont{Ekman}},
  \bibinfo{author}{\bibfnamefont{F.~A.} \bibnamefont{Asenjo}},
  \bibnamefont{and} \bibinfo{author}{\bibfnamefont{J.}~\bibnamefont{Zamanian}},
  \bibinfo{journal}{Phys. Rev.} \textbf{\bibinfo{volume}{E96}},
  \bibinfo{pages}{023207} (\bibinfo{year}{2017}), \eprint{1702.00722}.

\bibitem[{\citenamefont{Ekman et~al.}(2019)\citenamefont{Ekman, Al-Naseri,
  Zamanian, and Brodin}}]{Ekman:2019vrv}
\bibinfo{author}{\bibfnamefont{R.}~\bibnamefont{Ekman}},
  \bibinfo{author}{\bibfnamefont{H.}~\bibnamefont{Al-Naseri}},
  \bibinfo{author}{\bibfnamefont{J.}~\bibnamefont{Zamanian}}, \bibnamefont{and}
  \bibinfo{author}{\bibfnamefont{G.}~\bibnamefont{Brodin}},
  \bibinfo{journal}{Phys. Rev.} \textbf{\bibinfo{volume}{E100}},
  \bibinfo{pages}{023201} (\bibinfo{year}{2019}), \eprint{1904.08727}.

\bibitem[{\citenamefont{Bhadury
  et~al.}(2021{\natexlab{b}})\citenamefont{Bhadury, Florkowski, Jaiswal, Kumar,
  and Ryblewski}}]{Bhadury:2020puc}
\bibinfo{author}{\bibfnamefont{S.}~\bibnamefont{Bhadury}},
  \bibinfo{author}{\bibfnamefont{W.}~\bibnamefont{Florkowski}},
  \bibinfo{author}{\bibfnamefont{A.}~\bibnamefont{Jaiswal}},
  \bibinfo{author}{\bibfnamefont{A.}~\bibnamefont{Kumar}}, \bibnamefont{and}
  \bibinfo{author}{\bibfnamefont{R.}~\bibnamefont{Ryblewski}},
  \bibinfo{journal}{Phys. Lett. B} \textbf{\bibinfo{volume}{814}},
  \bibinfo{pages}{136096} (\bibinfo{year}{2021}{\natexlab{b}}),
  \eprint{2002.03937}.

\bibitem[{\citenamefont{Hu}(2021{\natexlab{b}})}]{Hu:2021plu}
\bibinfo{author}{\bibfnamefont{J.}~\bibnamefont{Hu}}
  (\bibinfo{year}{2021}{\natexlab{b}}), \eprint{2110.12339}.

\bibitem[{\citenamefont{Hehl}(1976)}]{Hehl:1976vr}
\bibinfo{author}{\bibfnamefont{F.~W.} \bibnamefont{Hehl}},
  \bibinfo{journal}{Rept. Math. Phys.} \textbf{\bibinfo{volume}{9}},
  \bibinfo{pages}{55} (\bibinfo{year}{1976}).

\bibitem[{\citenamefont{Speranza and Weickgenannt}(2021)}]{Speranza:2020ilk}
\bibinfo{author}{\bibfnamefont{E.}~\bibnamefont{Speranza}} \bibnamefont{and}
  \bibinfo{author}{\bibfnamefont{N.}~\bibnamefont{Weickgenannt}},
  \bibinfo{journal}{Eur. Phys. J. A} \textbf{\bibinfo{volume}{57}},
  \bibinfo{pages}{155} (\bibinfo{year}{2021}), \eprint{2007.00138}.

\bibitem[{\citenamefont{Becattini et~al.}(2013)\citenamefont{Becattini,
  Chandra, Del~Zanna, and Grossi}}]{Becattini:2013fla}
\bibinfo{author}{\bibfnamefont{F.}~\bibnamefont{Becattini}},
  \bibinfo{author}{\bibfnamefont{V.}~\bibnamefont{Chandra}},
  \bibinfo{author}{\bibfnamefont{L.}~\bibnamefont{Del~Zanna}},
  \bibnamefont{and} \bibinfo{author}{\bibfnamefont{E.}~\bibnamefont{Grossi}},
  \bibinfo{journal}{Annals Phys.} \textbf{\bibinfo{volume}{338}},
  \bibinfo{pages}{32} (\bibinfo{year}{2013}), \eprint{1303.3431}.

\bibitem[{\citenamefont{De~Groot et~al.}(1980)\citenamefont{De~Groot,
  Van~Leeuwen, and Van~Weert}}]{DeGroot:1980dk}
\bibinfo{author}{\bibfnamefont{S.~R.} \bibnamefont{De~Groot}},
  \bibinfo{author}{\bibfnamefont{W.~A.} \bibnamefont{Van~Leeuwen}},
  \bibnamefont{and} \bibinfo{author}{\bibfnamefont{C.~G.}
  \bibnamefont{Van~Weert}}, \emph{\bibinfo{title}{Relativistic Kinetic Theory.
  Principles and Applications}} (\bibinfo{publisher}{North-Holland},
  \bibinfo{year}{1980}).

\bibitem[{\citenamefont{Weickgenannt
  et~al.}(2022{\natexlab{b}})\citenamefont{Weickgenannt, Wagner, and
  Speranza}}]{Weickgenannt:2022jes}
\bibinfo{author}{\bibfnamefont{N.}~\bibnamefont{Weickgenannt}},
  \bibinfo{author}{\bibfnamefont{D.}~\bibnamefont{Wagner}}, \bibnamefont{and}
  \bibinfo{author}{\bibfnamefont{E.}~\bibnamefont{Speranza}},
  \bibinfo{journal}{Phys. Rev. D} \textbf{\bibinfo{volume}{105}},
  \bibinfo{pages}{116026} (\bibinfo{year}{2022}{\natexlab{b}}),
  \eprint{2204.01797}.

\end{thebibliography}

\end{document}